\documentclass[10pt, a4paper, english]{article} % document type and language

\usepackage{amsmath, amssymb, amsthm}
\usepackage{graphicx}
\def\al{\alpha}

\def\de{\delta}
\def\ga{\gamma}

\def\ep{\epsilon}

\def\te{\theta}
\def\la{\lambda}

\def\om{\omega}
\def\si{\sigma}

\def\De{\Delta}

\def\Te{\Theta}

 \def\calR{{\hbox{\cal R}}}

 \def\R{\mathbb{R}}

 \def\Z{\mathbb{Z}}

\def\ip{\hbox to4pt{\leaders\hrule height0.3pt\hfill}\vbox to8pt{\leaders\vrule width0.3pt\vfill}\kern 2pt}

\def\del{\partial}

\def\arr{\rightarrow}

\def\frac[#1/#2]{\hbox{$#1\over#2$}}
\def\Frac[#1/#2]{{#1\over#2}}
\def\({\left(}
\def\){\right)}
\def\[{\left[}
\def\]{\right]}
\def\^#1{{}^{#1}_{\>\cdot}}
\def\_#1{{}_{#1}^{\>\cdot}}
\def\Label=#1{{\buildrel {\hbox{\fiveSerif \ShowLabel{#1}}}\over =}}
\def\<{\kern -1pt}

\def\frame#1{\vbox{\hrule\hbox{\vrule\vbox{\kern2pt\hbox{\kern2pt#1\kern2pt}\kern2pt}\vrule}\hrule\kern-4pt}}

\def\Box to #1#2#3{\frame{\vtop{\hbox to #1{\hfill #2 \hfill}\hbox to #1{\hfill #3 \hfill}}}}

\def\ubal{\underline{\al}\kern1pt}
\def\obal{\overline{\al}\kern1pt}

\def\ubR{\underline{R}\kern1pt}
\def\obR{\overline{R}\kern1pt}
\def\ubom{\underline{\om}\kern1pt}
\def\obxi{\overline{\xi}\kern1pt}
\def\ubu{\underline{u}\kern1pt}
\def\ube{\underline{e}\kern1pt}
\def\obe{\overline{e}\kern1pt}

\def\ShowLabel#1{\ref{#1}}

\def\NewSection{\section}

\def\NewAppendix#1#2{\section*{Appendix #1: #2}}
\def\Acknowledgements{\section*{Acknowledgements}}
\def\ms{\medskip}

\def\eq#1{\begin{equation}#1\end{equation}}
\def\eqLabel#1#2{\begin{equation}#1\label{#2}\end{equation}}
\def\Cases#1{\begin{cases}#1\end{cases}}

\def\eqs#1{\begin{equation}\begin{aligned}#1\end{aligned}\end{equation}}
\def\eqsLabel#1#2{\begin{equation}\begin{aligned}#1\end{aligned}\label{#2}\end{equation}}

\def\Note{\begin{quote}\small}
\def\EndNote{\end{quote}}

\newtheorem{defn}{Definition}[section]
\def\Definition{\begin{defn}}
\def\EndDefinition{\end{defn}}

\def\Figure[#1]#2{\begin{figure}[htbp] %  figure placement: here, top, bottom, or page
   \centering
   \includegraphics[#1]{#2} }

\def\EndFigure{\end{figure}}

\title{Relativistic GPS in 3-dimensions}

\author{S.Carloni$^{a, b}$, L.Fatibene$^{c,d, e}$, M.Ferraris$^c$, R.G.McLenaghan$^e$, A.Orizzonte$^{ c,d}$\\
\\\it\small
$^a$ DIME Sez. Metodi e Modelli Matematici, Università di Genova, (Italy).\\\it\small
$^b$ Institute of Theoretical Physics
Faculty of Mathematics and Physics, Charles University (Czech Republic)\\\it\small
$^c$ Department of Mathematics, University of Torino (Italy)\\\it\small
$^d$ INFN - Sezione Torino (Italy)\\\it\small
$^e$ Department of Applied Mathematics, University of Waterloo (ON-Canada)}

\begin{document}
\maketitle

\begin{abstract}
We extend to {three dimensions} the proposal of a completely relativistic positioning system (rPS).
The system does not rely on approximations, in fact, it  works at a few Schwarzschild radii from a black hole, and it does not rely on Newtonian physics
or special relativity. 
Since general relativity (GR) claims to be our fundamental framework to describe classical physics, it must provide tools to bootstrap physics within the theory itself, without relying on previous approximated frameworks.

The rPS is able to {\it self-diagnose}, that is, it detects deviations from assumptions about the gravitational field and consequently stops operations; in addition it is {\it robust}, i.e., it is able to autonomously restore operations when assumptions are restored.

From a more general viewpoint, the rPS is equivalent to geodesy in spacetime, which establishes a (conventional) coordinate system on a surface by means of measurements within the surface itself, as well as allowing it to extract information about the intrinsic geometry of the same surface. In other words, the positioning system is potentially able to extract information about the gravitational field (which in fact is identified with the geometry of spacetime) in addition to the gravitational theory, which describes its dynamics. Thus, it becomes a framework {within which} one can operationally distinguish different theories of gravitation. 
\end{abstract}

\NewSection{Introduction}

In a previous paper \cite{rPS} we proposed a setting for a toy model relativistic positioning system (rPS), which is able to determine the position of a user (in spacetime), together with the orbital parameters of the constellation of satellites (up to isometries) as well as to check the assumptions made about the gravitational field and other forces acting on {the} satellites.
The idea is not new since there are many similar though not identical proposals; (see \cite{Coll1}, \cite{Blago}, \cite{Coll2}, \cite{Coll3}, \cite{Coll4}, \cite{Coll5}, \cite{Tarantola}, \cite{Rey}).

%To be honest, 
The whole project 1s based on a rather naive initial question, namely, {\it why should one need a {\it ground control}, also named {\it control segment} in the GPS literature, when, in principle, one could use the GPS infrastructure itself to determine the position of satellites' constellation?}
We also aim to investigate a system which {can} function in a strong field regime (namely something which {can} work well at a few multiples of the Schwarzschild radius of a compact object) producing, in principle, exact results.
Moreover, we {aim} for a fully relativistic system that {does} not resort in any way to Newtonian or special relativistic physics.
We {desire} a system which {does} not assume structures fixed on spacetime other than a given gravitational field, {and requires} no external synchronizations, no external validation, {and} no calibration. 

Of course,  in \cite{rPS}, we did some theoretical simplifications:
first of all, we assumed satellites to carry proper clocks and to be free falling and, secondly, we restricted to 1+1 dimensions, 
namely a system of {one} spatial dimension. As a result, the system presented was a toy model, a proof of concept for a system that needed further development. 

In 1+1 dimension, unlike in dimension 2+1, one has no orbits {\it around} a compact object, all satellites either fly away or into the central compact objects, in the second case soon disappearing {over} an event horizon. 
The problem is also greatly simplified in view of the fact that any light ray emitted, sooner or later reaches any satellite moving on {the same} side.
In 1+1 there is no control problem to be solved, namely one has only to select the initial direction to hit the satellite. 
The analogous problem in 2+1 is much more complicated since the satellite is moving and the gravitational field, at the same time, is bending the light ray trajectories.

%cope with 
Working in dimension 2+1, we show {that} the toy model can be extended and that these problems do not impact in a fundamental way on an rPS.
At {a minimum}, we collect mathematical methods to treat the problem in GR, in all its glory.
We 
%here 
present a rPS which does not make any weak field approximation and does not require an external synchronization or calibration,
in other words, it does not need a control segment.
It determines all orbital parameters of the space segment, namely the satellites and the clocks they carry 
%around, 
under the assumption of a static, rotationally symmetric gravitational field. 
The system is able to diagnose deviations from {prior assumptions} ({e.g.,} the free falling of satellites, or about the gravitational field), and notify  the users to prevent ill positioning. 
The system is also able to autonomously restore operations when these anomalies disappear.
In principle, it can also model deviations perturbatively by regarding some perturbations as part of the variables to fit together with the orbital parameters of the space segment.

The procedure is not, in principle, different from the one used in geodesy: one measures relative distances (or angles) between a number of points fixed on a surface (e.g., the surface of the Earth)
%and then 
from which measurements one can extract the intrinsic geometry of the surface, e.g., determine the curvature of the Earth's surface from intrinsic measurements \cite{Gauss1845}.
%on its surface. 
If one of the points is the position of the user and the surface is replaced by spacetime, that is exactly what an rPS does; it determines the position of the user with respect to the other points in the constellation as well as the intrinsic geometry of the surface (which in the case of spacetime is the gravitational field).

A similar problem has been studied {by other authors from} a more theoretical viewpoint; (see \cite{Bini}, \cite{Rovelli1}).
%As a matter of 
In fact, a rPS is  a recognized way of measuring the gravitational field and defining observables in GR, 
which in vacuum is well known to have no non-trivial observables.

Moreover, {by} claiming GR to be a fundamental theory, specifically, our framework to understand classical physics, one {expects to able} to provide, within GR, a way of bootstrap fundamental experiments to describe {classical phenomena}. The first step is to be able to describe the motion of test particles in a spacetime neighbourhood without relying on a Newtonian (or SR) scheme in which coordinates are endowed with a direct meaning of distances, and one selects a class of observers (namely inertial observers) which gives us a description of physics which defines structures on spacetime (namely, affine structure or absolute time and space) which we know to be unphysical.

A relativistic positioning system is a necessary step that enables us to describe the motion of test particles around an observer, that is,
a first step {towards the extraction of the} laws of motion without {reliance} on {the} Newtonian ones that we know are only approximately true.
%reliable.
%\notaS{It should be remarked at this point that the procedures shown in the following have only a ``proof of concept'' and they are not optimised for, e.g., calculation speed.}

\ms
{The paper is organized as follows:} in Section 2, we review how one may describe analytically a satellite constellation.
In Section 3, we consider the description of light rays, in particular how to find a ray emitted from a satellite {which hits a target event}.

In Appendix A, we discuss the structure of the light {conoid}, i.e., the ``light conoid'' in spacetime (as opposed to the light cone in the tangent space).
This study allows us to divide light rays into classes depending on whether they hit the BH or are simply scattered, or whether they go around the BH clockwise or counterclockwise and, if needed, to divide each ray into branches which can be suitably parameterized.
We note that each class corresponds to a region on the light {conoid} which also exactly accounts for all details of strong lensing \cite{Schneider}.

In Appendix B, we present a symplectic framework to describe light rays which is equivalent to the Lagrangian formalism used in Section 3.
However, it provides a framework to solve in general the control problem of finding a solution of a Hamiltonian system passing through an initial and a final position.

In Section 4, we set up a rPS and discuss how one can parameterize the signals received at a target event $p_0$ on a satellite $\chi_0$.
In Section 5, we give an example of the simulation of the model in the case of a rPS around a BH in the strong regime.
In Section 6, we similarly describe the parameters of a rPS around the Earth, with orbital parameters similar to the actual NAVSTAR-GPS.
These two cases present different challenges, and although they are particular cases of a general framework, they can be optimized in different directions.
Finally, we briefly discuss conclusions and perspectives in Section 7.

\NewSection{Satellite constellation}

We have to deal with two problems:~describing the orbits of satellites $\chi_i$ and the light rays exchanged between them.
We start with the first problem, since the second will {be a modification of} it.
We consider the Schwarzschild metric in dimension 2+1 which,  with respect to the standard coordinate system $(t, r, \te)$, has the form
\eq{
g= -A(r) c^2 dt^2 + B(r) dr^2 + r^2 d\theta^2,
}
where $A(r)= 1 -\frac[\al/r]$, $B=\frac[1/A]$, $\alpha$ is {a constant related to the mass of the central object}, {and $c$ denotes  the speed of light in the limit $r\arr \infty$}. This is the most general stationary, spherically symmetric solution of the vacuum Einstein equations.
{Here,} unlike  the 1+1 case, we have real {\it bounded orbits}, which we can {extend to} the 3+1 situation, {by} restricting the constellation (and the user) to {be on} the equatorial plane.

The {basic} idea is to {avoid} approximations and {instead,} use the Lagrangian formalism  to reduce the problem to Weierstrass equations {which can} be integrated by semi-analytic methods.
We use the parameter invariant Lagrangian for describing timelike geodesics, since 
it is well known that it describes {the} worldlines of test particles and, {since it is invariant with respect to parameterizations,} {it} can also be  used {later} for light rays on which, of course, we cannot define proper time. 
The invariant arc-length associated with the Lagrangian defines the geodesic distance between two points in spacetime. The standard definition of the \emph{distance-function} {$\Gamma$} is the square of the geodesic distance with the sign according to whether the non-light-like geodesic is time-like or space-like.  (See \cite{Hadamard}, \cite{Ruse1}, \cite{Ruse2}, \cite{Ruse3}, \cite{Synge1}, \cite{Yano}, \cite{Schouten}, \cite{Friedlander}, \cite{Benenti}). Synge \cite{Synge2} calls { $\Omega = \frac[1/2] \Gamma$} the world function.

{He argues} that one can deduce from it the geometric structure of spacetime, which makes sense since the geometric structure is {determined by} the gravitational field {via Eienstein's equations}, and the gravitational field, as any other field, is encoded and observable only through the effects it has on test particles, the motion of which is described by the invariant Lagrangian.

{We can choose any parameterization gauge, but} here we shall initially use  {coordinate} time $t$, not the proper time; we shall try to highlight why later on.
Hence the Lagrangian {for timelike worldlines} is
\eqLabel{
\tilde L=  \sqrt{A c^2 (t')^2 - \Frac[(r')^2/A] - r^2 (\theta')^2} \> ds
}{piL}
where primes denote derivatives with respect to the arbitrary parameter $s$.
{Choosing $t$ as the parameter we} obtain
\eqLabel{
L=  \sqrt{Ac^2 - \frac[\dot r^2 / A] - r^2 \dot\theta^2} \> dt,
}{Lag}
where the dots denote derivatives with respect to $t$.

{Since the Lagrangian $L$ does not depend explicitly on $t$ or on $\theta$ we have the following two first integrals:}
\eqLabel{
{c^2}\ep=-\Frac[{c^2}A/\sqrt{Ac^2 - \frac[\dot r^2 / A] - r^2 \dot\theta^2}],
\qquad\qquad
k=-\Frac[r^2 \dot \theta/\sqrt{Ac^2 - \frac[\dot r^2/A] - r^2 \dot\theta^2}],
}{FirstIntegrals}
{where $\ep$ and $\theta$ are constant.}
{Solving for the Lagrangian velocities we get}
\eq{
\dot r^2 =\Frac[A^2/r^2\ep^2]((\ep^2c^2-A)r^2 -A k^2),
\qquad\qquad
\dot \theta^2 =\Frac[k^2A^2/r^4 \ep^2].
}
These can be cast into the form of Weierstrass equations
\eqs{
\(\Frac[dr/d t]\)^2=& \Frac[A^2/ r^2\ep^2]((\ep^2c^2-A)r^2 -A k^2) =: \Phi(r; \ep, k)  \cr
\(\Frac[dr/d\theta]\)^2 =&\Frac[r^2/ k^2]  ((\ep^2c^2-A)r^2 -A k^2)= :\Psi(r; \ep, k)
}
{The functions $\Phi$ and $\Psi$ are called {\it Weierstrass functions} or also {\it effective potentials}}.
The value of $(\ep, k)$ parameterizes the solutions since they can be obtained from initial conditions. Both of them diverge to $-\infty$ for light rays.
The motion of the system {takes place} where $\Phi\ge 0$, which is called the {\it allowed region}.

%\Note
{We} remark that $\ep$ and $k$ are not exactly physical energy and angular momentum (as dimensional analysis shows directly). 
If one wants to map first integrals to physical quantities, one can compare them to what happens in the case of {Keplerian motions}.
%\EndNote

The next step will be to {set} $A= 1- \frac[\alpha/r]$, i.e.~{the} Schwarzschild {solution} rather than working on a generic static, spherically symmetric metric.

%\Note
If one {uses} $A= 1- \frac[\alpha/r] + \lambda r^2$, one would have AdS-Schwarzschild. This {case will} be left for future investigation, 
e.g., to determine how long we should observe the system to detect a given non-zero $\la$.
%\EndNote

By {the substitution} $A= 1- \frac[\alpha/r]$ 
%(and $\alpha=1$, i.e.~we measure distances as multiples of the Schwarzschild radius)
into {the} definitions, {the} Weierstrass functions {take the following form:}
\eqsLabel{
\Phi(r; \ep,k) =& \Frac[(r-\al)^2/r^5 \ep^2] ((\ep^2c^2-1)r^3+ \al r^2- k^2 r +k^2\al )\cr
\Psi(r; \ep, k) =&\Frac[r/k^2] ((\ep^2c^2-1)r^3+ r^2\al- k^2 r +k^2\al)
}{WEJ}
\begin{figure}[htbp] %  figure placement: here, top, bottom, or page
   \centering
   \includegraphics[width=10cm]{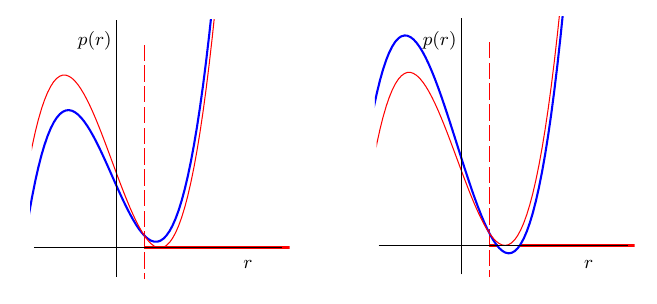} 
   \caption{{\it we show the qualitative allowed region of polynomial $p(r)$ for massive particles and $\ep^2c^2-1>0$. The dashed line is at {the} Schwarzschild radius.
   The red thin line is the limiting case with an asymptotic circular orbit.\\
   A thick blue line on the left {shows} one allowed region; the material point falls into the horizon.\\
   The thick blue line on the right {shows} two allowed regions; the material point either falls into the horizon or {is scattered} to infinity.}}
   \label{figPlotWm1}
\end{figure}

The allowed regions are {determined} by the sign of the polynomial %x\notaR{Suggest that $p(r)$ be displayed with a number} 
\eq{
p(r)=(\ep^2c^2-1)r^3+ \al r^2- k^2 r +k^2\al
}
{we shall} focus on the external region $r>\al$.
{For} bounded orbits, there should be an allowed region $[r_-, r_+]$ in the external region, which {implies that} $r=r_\pm>\al$ be a simple {root} of the polynomial, $p(r_\pm)=0$.
{Since this} is a polynomial of degree 3,  it has at least {one and at most three real roots}. 
{Note} that $p(\al)= \al^3c^2\ep^2\ge 0$.
{As $r \rightarrow \infty $} , $p(r)$ diverges to infinity with a sign {given} by the leading coefficient $\ep^2c^2-1$.

If one has $\ep^2c^2-1>0$ (see Fig.\ref{figPlotWm1}), the polynomial $p(r)$ has either 0 or 2 roots in the external region.
If there are zero roots, the whole exterior region is allowed, and the test particle either escapes to infinity or falls into the BH.
If there are two roots, one has two allowed regions, one near the horizon where the test particle falls and one near infinity where the test particle escapes to infinity. In both cases {there are} no bounded orbits.
If $\ep^2c^2-1=0$, the leading term becomes $r^2$, {which implies} either 0 or 2 solutions, {and} in both cases no bounded orbits.

\begin{figure}[htbp] %  figure placement: here, top, bottom, or page
   \centering
   \includegraphics[width=10cm]{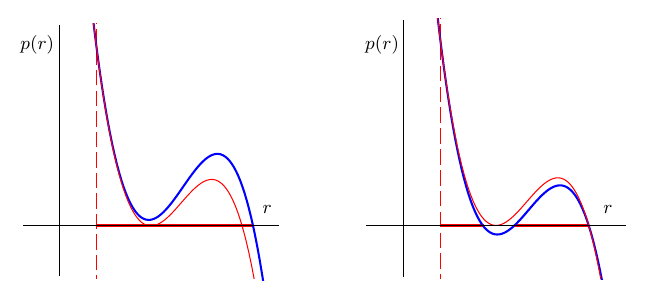} 
   \caption{{\it Qualitative allowed region of polynomial $p(r)$ for massive particles and $\ep^2c^2-1<0$. {The} dashed line is at {the} Schwarzschild radius.
   The thin red line is the limiting case with an asymptotic circular orbit.\\
   {The} thick blue line on the left {shows} one allowed region; the material point falls into the horizon.\\
   {The} thick blue line on the right {shows} two allowed regions; the material point either falls into the horizon or goes along a bounded orbit.}}
   \label{figPlotW2}
\end{figure}

If $\ep^2c^2-1<0$ (see Fig.\ref{figPlotW2}), the polynomial has either 1 or 3 roots in the external region.
{Only} one solution means an allowed region near the horizon, {across which} all particles fall. {which again implies} no bounded orbits.
We are left with the last option (3 real roots) which yields two allowed regions: one close to the horizon {across which} test particles fall and {the other} which describes bounded orbits.
{As a consequence of the above analysis we are lead to} consider {the case of} three roots {where} we {want to recast the polynomial $p(r)$ in the form}:
\eq{
(\ep^2c^2 -1)(r-r_0)(r-r_1)(r-r_2)
}
{By comparison}, we {obtain}  three equations for the unknowns $(\ep^2, k^2, r_0, r_1, r_2)$ {which can be solved for} $(\ep^2, k^2, r_0)$ as functions of $(r_1, r_2)$. {Since expressing the} roots as functions of $(\ep^2, k^2)$ is difficult, we prefer to write $(\ep^2, k^2)$ (and $r_0$) as a function of the other two roots {which form} the boundary of the allowed region $[r_1=r_-, r_2=r_+]$ associated to the bounded orbit. In this way, we are parameterizing solutions by perihelion (perigee) $r_-$ and aphelion (apogee) $r_+$, instead of $(\ep^2, k^2)$ or initial conditions.
This is convenient since $r_\pm$ are also integration limits of improper integrals {one which may be used to determine} the orbital period. {In this case the appropriate}       
Weierstrass functions in terms of $(r_-, r_+)$ are {given by}
\eqs{
\Phi(r; r_-, r_+) =&\Frac[c^2 \al (r-\al)^2 (r_+-r)(r-r_-)/ (r_+-\al)(r_--\al)(r_++r_-) r^5]\((r_+r_--\al(r_++r_-))r - \al r_+r_-\)\cr
\Psi(r; r_-, r_+) =&\Frac[r(r_+-r)(r-r_-)/ r_+^2 r_-^2]\((r_+r_--\al(r_++r_-))r - \al r_+r_-\)\cr
}

We can {utilize} {any computer algebra program} to {evaluate} the integral, namely
\eq{
t(r; r_+, r_-)= \int_{r_-}^r \Frac[dR/\sqrt{\Phi(R; r_+, r_-)}]
}
{We} stress that Maple or Mathematica can {evaluate} this integral analytically precisely because the polynomial $p(r)$ is factorized, {which results from the fact that the orbits are parameterized with respect to their aphelion and perihelion values $r_\pm$.}

{We} also remark that even though we started with a parameterization in terms of the relative time $t$, 
we eventually get orbits parameterized by $r$. 
We argue that we do not need or should try to resist this.
After all, we are interested in the orbits, and the parameterization  is just a {means} to describe them.

{In order to follow the satellites for as many orbits as we wish, the branches which describe an orbit must be glued together.}   {Thus we require} a standardized way of denoting {the} different branches.

Hence, we start at the perihelion $r=r_-$ at $t=t_0$. {On} branch 0, $r$ {increases} to aphelion $r_+$ in time
$t(r; b=0)=  t_0 + t(r;  r_+, r_-)$ {which} it reaches {at} time $t_1= t_0+T$, where we set $T:= t(r_+; r_+, r_-)$ for half of the $t$-period.
 Branch 1 starts at aphelion $r_+$ at $t_1$ and $r$ {decreases} to $r_-$.  {Branch 1 is described as}
$t(r; b=1)=t_1+ T - t(r; r_+, r_-)$. It {reaches} $r=r_-$ at {time} $t_2=t_0+ 2T$.
{Repeating the process for branch 2 we have} $r$ increasing {from $r=r_-$} {to $r=r_+$ in time} $t(r; b=2)= t_2 + t(r; r_+, r_-)$.
It ends the branch {at time} $t_3= t_2+T= t_0+ 3T$.
On branch 3, $r$ decreases from $r_+$ to $r_-$ with $t(r; b=3)= t_3+T-t(r; r_+, r_-)$. It ends {when} $r=r_-$ and $t_4=t_3+T= t_0+ 4T$. {This procedure may be continued}.
We also have branch -1 in which $r$ decreases from $r_+$ to $r_-$ where branch 0 starts.
During the branch $t(r; b=-1)= t_0-t(r; r_+, r_-)$. Accordingly, branch -1 starts at $r_+$ and $t_{-1}= t_0-T$.
Similarly, during branch -2,  $r$ increases from $r_-$ to $r_+$ with $t(r; b=-2)= t_0-2T+t(r; r_+, r_-)$. And so on.

At the same time, we  {can determine the angle $\theta$ by}
\eq{
\theta(r; \te_0, r_+, r_-)=\te_0\pm \int_{r_-}^{r} \Frac[dR/\sqrt{\Psi(R; r_+, r_-)}]
}
{For definiteness} we shall {consider} counterclockwise orbits, which correspond to {the} $+$ sign on outgoing (even) branches ($\te$ increasing with $r$)  
and {the} $-$ on ingoing (odd) branches ($\te$ {increases as} $r$ decreases).
During a branch, the angle increases {by the amount}
\eq{
\Theta=\int_{r_-}^{r_+} \Frac[dR/\sqrt{\Psi(R; r_+, r_-)}]
}
If this quantity is $\pi$ (as in Keplerian motions), the orbit is closed. {The} deviation $\de=2(\Te-\pi)$ measures {the} precession in radians per orbit.

For {the} proper time, we have
\eq{
d\tau =c^{-1} \sqrt{A c^2(t')^2 - \frac[1/A] (r')^2 - r^2 (\theta')^2} \> ds
 =c^{-1} \sqrt{ \frac[Ac^2/\Phi] - \frac[1/A]  - \frac[r^2/\Psi] } \> dr
}
{Thus},  we can define
\eq{
\tau(r; \tau_0, r_+, r_-)= \tau_0\pm c^{-1} \int_{r_-}^{r}  \sqrt{ \frac[Ac^2/\Phi] - \frac[1/A]  - \frac[R^2/\Psi] } \> dR
}
This {result} allows us to keep track of proper time along branches so that everything is locally parameterized in $r$.
{By this means}, we can {compute} the orbit of a satellite in spacetime  back and forth for {as} many branches {as} we need.
{In addition}, the whole orbit of a satellite is determined by $(t_0, \te_0, r_+, r_-)$. 
The proper time is known once we set $\tau_0$, namely the proper time at perihelion on the branch $b=0$.
See the Maple drawing of the orbit in Figure 3.

\begin{figure}[htbp] %  figure placement: here, top, bottom, or page
   \centering
   \includegraphics[width=5cm]{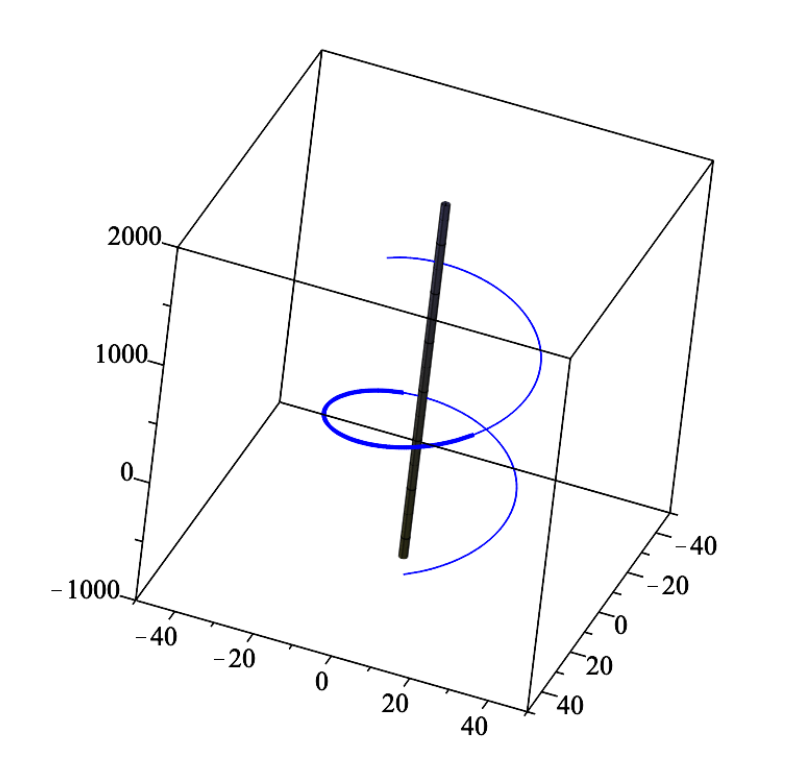}
   \qquad\qquad
   \includegraphics[width=5cm]{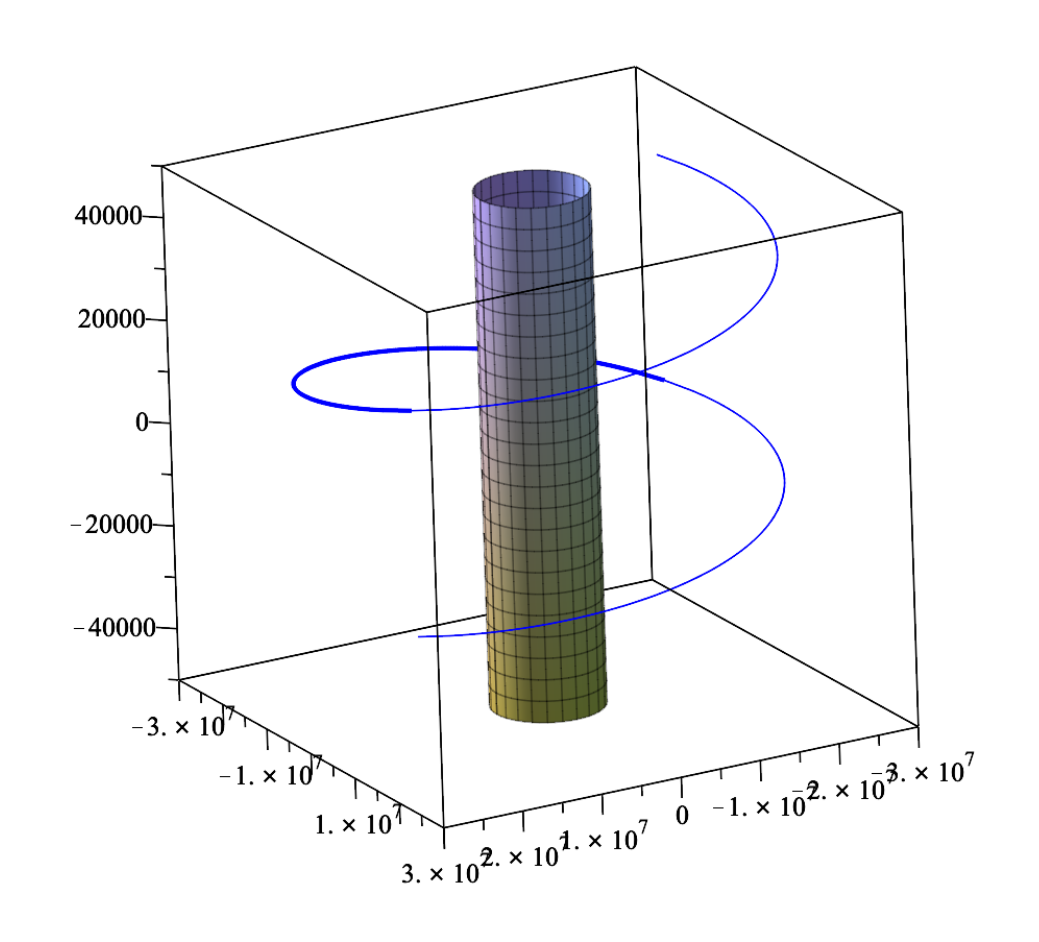} 
\centerline{\it\hskip 1cm {\bf BH} \hskip 5cm {\bf Earth}}
\caption{\small\it{\bf BH}.
We set $\al=1$ and $c=1$. Orbital parameters are $r_-=20\al$, $r_+=28\al$, $t_0=10c^{-1}\al$,  $\te_0=\frac[7\pi/6]$, 
and $\tau_0=-c^{-1}\al$.
The continuous line is the (exact analytical) satellite orbit for the branches -1, 0, and 1.
The thick part is branch 0. 
The cylinder at the center is the Schwarzschild event horizon at $\al=1$.
The orbit {precesses by} $\de=2\Te-2\pi= 0.4477$ (radians per orbit).
{The precession can be computed to any desired precision} (e.g.~in this case, we have 
$\de= 0.4476574054738341735420832219320682$ which is clearly more than we can hope to measure).
The orbital period is $2T=1117.0454$ (in units fixed by setting $c=1$) in relative time,
$2\tau= 1081.5289$ in satellite proper time.
Hence, we see that the satellite clock slows down (with respect to coordinate time) by a factor of $0.9682$ 
as an effect of the gravitational field of the BH.
Again we can {compute these quantities to any desired}  precision. \\
\\
{\bf Earth}.
Here we use SI {units}. We set Earth parameters $\al =0.008870355 \>m$ and $c= 299792458 \>ms^{-1}$.
Orbital parameters are $r_-=24289000\>m$, $r_+=25089000\>m$, $t_0=-1600\>s$, and $\te_0= -\frac[\pi/6]$ $\tau_0=-1600\>s$, which are comparable to those of a GPS satellite.
The continuous line is the (exact analytical) satellite orbit for the branches from -1 to 1.
The thick part is branch 0. 
The cylinder at the center of the orbit is the worldsheet of Earth's surface at $3189 \>km$.
The {orbital precession is}  $\de=2\Te-2\pi\simeq  3.4\cdot 10^{-9}$ (radians per orbit).
The orbital period is $2T=38606.364226049\> s$ (i.e.~about $10.7h$, which is comparable with what is expected for  NAVSTAR-GPS satellites) in Earth's relative time,
$2\tau= 38606.364215646\>s$ in satellite proper time.
The slow down of proper time cannot be {observed} at this level of precision. However, the absolute slowing down can be computed to be
$2T-2\tau =1.040 \times 10^{-5}\> s$.
Hence, we see that in these conditions {the} satellite precession is small, and the satellite clock slows down by about $10\> \mu s$ per orbit.}
\end{figure}

It is not difficult to add further satellites (two more, in fact) on similar orbits. 
For the sake of simplicity, we select the orbital parameters so that the orbits never cross, each satellite being in a ring disjoint from the others,  starting from $\chi_0$ in the smaller ring, $\chi_1$ being in the middle, and $\chi_2$ being {the furthest out}. 
In this way, we have a constellation of 3 satellites, which (at least when they do not eclipse each other) is a minimal constellation as discussed in \cite{rPS}.

Before {considering} light rays, {we remark} that if we {had}  {utilized} the quadratic Lagrangian, which {is known to provide geodesics}  parameterized by proper time, we would, {accordingly}, be looking for 
solutions {in the form} $\tau(r; r_\pm)$.

In {view of our solution, we can now guess how to produce the solution parameterized by proper time: we should have to write}  $\tau(r(t); r_\pm) $,
where the function $r(t)$ would need to be obtained by solving {our solution} $t=t(r; r_\pm)$.
In our case, though, the function $t(r; r_\pm)$ is explicitly computed, and we see it is a (rather complicated) combination of elliptic functions. 
The inverse cannot be easily expressed in terms of special functions,
as it would not be for the solutions of the quadratic Lagrangian. 
For this reason, it is convenient to use relative time {rather than} proper time (and eventually the coordinate $r$), even if proper time may have a better {\it physical meaning}.

\NewSection{Light rays description}

{We now turn to the determination of the orbits of light rays in the Schwarzschild spacetime. This task is relatively easy, since we are able to use the formalism developed in the previous section to determine the satellite orbits. This is possible because for the earlier calculation we used the parameter invariant Lagrangian (\ShowLabel{piL}) instead of the more usual quadratic one based on proper time parameterization.} 

It is well known (see \cite{EPS}, \cite{OurEPS1}, \cite{OurEPS2}) that light rays can be {approached arbitrarily closely} by {the orbits of} material {particles}. Thus, we can obtain a description of light rays as the limit of a sequence of {orbits} closer and closer to the light {conoid}. {Regarding terminology} , we call {\it light cone} {with vertex $p_0$} the subset of the tangent space {at $p_0$ defined by the lightlike lines through $p_0$} , while we call {\it light {conoid}} {with vertex $p_0$}, 
the envelope of lightlike worldlines {through $p_0$} as a subset of spacetime. 
{An event $p$ belongs to the light {conoid} with vertex $p_0$} iff there exists a light ray {connecting} $p$ {to} $p_0$.

Both the first integrals $\ep$ and $k$ diverge for light rays, as one can see from equations (\ShowLabel{FirstIntegrals}).
However, the quotient $K= \frac[k/\ep]$ is finite.
As a consequence, {the light rays are} obtained by replacing $k=K\ep$ in the Weierstrass equations {(\ShowLabel{WEJ})}, then letting $\ep\arr -\infty$. 
{This procedure} gives {the} Weierstrass equations directly for light rays as
\eqsLabel{ 
\(\Frac[dr/d t]\)^2=& \Frac[(r - \al)^2/r^5] (c^2 r^3-  K^2 r + K^2\al) =: \Phi(r; K)
\cr
\(\Frac[dr/d\theta]\)^2 =&\Frac[r/K^2] (c^2r^3-  K^2 r + K^2\al) = :\Psi(r; K)
}{LightRaysWeierstrass}

\begin{figure}[htbp] %  figure placement: here, top, bottom, or page
   \centering
   \includegraphics[width=10cm]{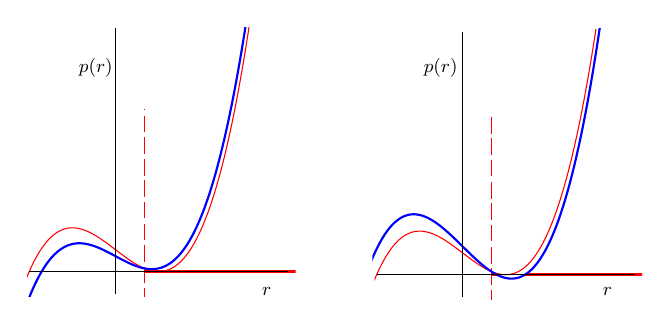} 
   \caption{{\it Qualitative allowed region for the polynomial $p(r)$ for light rays.The dashed line is at the Schwarzschild radius.
   The red thin line is the limiting case of an asymptotic circular orbit at $r=\frac[3/2]\al$.\\
   The thick blue line on the left illustrates one possible allowed region where the light ray crosses  the horizon.\\
   The thick blue line on the right illustrates two allowed regions: the light ray either crosses the horizon or it scatters to infinity.}}
   \label{figPlotWl}
\end{figure}

{We can now} use the techniques {that we employ} for the {satellites} with one less parameter. 
The real roots of the Weierstrass functions are governed by the polynomial $p(r)=c^2r^3-  K^2 r + K^2\al$. They can be $1$ or $3$ real roots (see Fig.\ref{figPlotWl}).

The case of 1 real root {occurs when} $0\le K^2\le \frac[27/4]\al^2c^2=: K^2_{cr}$, {in this case the root $r_0$ satisfies the inequality} $r_0\le -3\al$.
The outer region $(\al, +\infty)$ is {entirely} allowed, {which means} that a light ray either falls into the BH or escapes to infinity.
These light rays, called {\it in-falling rays}, {form a} single branch, either {\it ingoing} or {\it outgoing}.
The worldline of an in-falling light ray in spacetime is {defined by the} parameterized curve 
\eqLabel{
t(r; t_0, K) = t_0 + e \int_{r_0}^r \Frac[dR/\sqrt{\Phi(R; K)}]
\qquad
\te(r; \te_0, K) = \te_0 + \si e \int_{r_0}^r \Frac[dR/\sqrt{\Psi(R; K)}]
}{SolutionInfallingRays}
where $e=\pm 1$ {determines whether} the branch is {\it ingoing} ($e=-1$) or {\it outgoing} {($e=+1$)}, and $\si=\pm 1$ denotes {whether} the light ray goes around the BH {\it clockwise} ($\si=-1$) or {\it counterclockwise} {($\si=+1$)}.

In the boundary case $K^2= K^2_{cr}$, one has an asymptotic goal for $r=r_{cr}=\frac[3/2]\al$ which corresponds to a double root of $p(r)$.
That is a single ray, we ignore it since the integrals diverge in {this} case.

In the case of 3 distinct real solutions (i.e.~$K^2> K^2_{cr}$), it is convenient to {write} the polynomial {in factored form:} $p(r)=c^2r^3-  K^2 r + K^2\al = c^2(r-r_1)(r-r_m)(r+r_1+r_m)$,  with $0< r_1 \le r_m$ where $r_m$ is the minimal approach radius of the light ray.
In this case, we can express $K^2$ and $r_1$ as a function of $r_m$, which becomes the only parameter. {Thus} we can express the Weierstrass functions as $\Phi(r; r_m)$ and $\Psi(r; r_m)$.
{It follows that} there are two allowed regions: 
one is confined in the region $\al \le r\le r_1<\frac[3/2]\al$, the other is $r\in [r_m, \infty)$ which corresponds to a light ray that arrives from and escapes to spatial infinity.
Accordingly, such rays are called {{\it scattered rays}}.

A {scattered} light ray is entirely determined by $r_m$. {It consists} of two branches, an {\it ingoing} and an {\it outgoing} branch. 
By writing {the} Weierstrass functions in terms of $r_m$ we can use Maple or Mathematica to integrate the equations analytically 
\eqsLabel{
t(r; t_0, r_m) =& t_0 + e \int_{r_0}^r \Frac[dR/\sqrt{\Phi(R; r_m)}]\\
\te(r; \te_0, r_m) =& \te_0 + \si e \int_{r_0}^r \Frac[dR/\sqrt{\Psi(R; r_m)}]
}{SolutionScatteringRays}
in terms of elliptic functions as we did for {the} satellites.

{It follows that} we can fix an event $p_0$, choose a value for $r_m$, and draw exactly one light ray through $p_0$. 
{The} light ray reaches a minimum $r=r_m$ {at some point on its worldline} in the past or in the future of $p_0$  and escapes again to spatial infinity.
{In} Appendix $A$ {we use Maple to} draw explicitly the exact past {conoid} through an event $p_0$.
The surface we obtain accounts for all multiple images {of a particle} {that an observer at $p_0$} can {observe} around the BH (namely, its strong lensing {effect}\cite{Schneider}).
We also see that the surface is regular in a neighbourhood of $p_0$ except at $p_0$ itself, namely, {it is locally} a {\it conoid}. 
Further away from the vertex, the surface becomes multi-sheeted and self-intersecting, which is precisely what determines the strong lensing {effect}.
{An} important {feature of} the light {conoid} we obtain is that we {are able to} draw it {\it exactly} and {\it analytically}. 

%\Note
To summarize the situation, for each satellite on a bounded orbit, we have $(\te_0, t_0)$ the initial position given at the perihelion $r_-$ on branch 0,
as well as $(r_-, r_+)$ to identify the orbit. Since the satellite carries a proper clock, we also need an initial {value} $\tau_0$ for the proper time.
Then one can compute the half-periods $T$, $\Te$, and $\tau$. 
Each orbit then splits into a countable number of branches parameterized by $b\in\Z$, alternating {between} ingoing (odd $b$) and outgoing (even $b$) half-orbits.

For a light ray, we have in-falling (one for each sub-critical $K^2$) and {scattered} (one for each $r_m>\frac[3/2]\al$) {ray}.
In-falling light rays can be ingoing or outgoing and clockwise or counterclockwise.
Each in-falling light ray is made of a single branch.

On the {other hand}, each scattered light ray {consists} of two branches, one ingoing and one outgoing. They are classified by the minimal approach parameter $r_m$, {and} by whether the target is on the ingoing or outgoing branch and whether they go around the BH clockwise or counterclockwise.
%\EndNote

However, for positioning, we need instead to solve a control problem.
Imagine we fix an event $p$ (possibly, but not necessarily, on a satellite) and consider a(nother) satellite (namely, a {\it generic} timelike worldline $(t_s(r), r, \te_s(r))$, $s$ standing for {\it satellite} (or {\it source} of the message). 
We need to find a light ray through $p$ and some point $p_i$ of the source worldline corresponding to some value of the parameter $r=r_i$.
If {this is possible}, we can determine the proper time $\tau_i$ of the clock on the satellite at emission.

%\Note
It is {now} clear how to {do} this.
The satellite worldline is described by 
\eq{
(r=r_s, \te_s(r_s; \te_0, r_\pm), t_s(r_s; t_0, r_\pm))
} 
while the light ray through the target $p=(r_t, \te_t, t_t)$ is described by
\eq{
(r, \te_l(r; t_t, \te_t, r_m), t_l(r; r_t, t_t, r_m))
}
(or, equivalently, {by} $(r, \te_l(r; r_t, \te_t, K^2), t_l(r; r_t, t_t, K^2))$ for an in-falling ray). 
{Thus} a point $r_s$  on both the ray and the source worldline {must obey} the equations
\eq{
\te_l(r_s; t_t, \te_t, r_m) = \te_s(r_s; \te_0, r_\pm)+ 2n\pi
 \qquad
 t_l(r_s; r_t, t_t, r_m) = t_s(r_s; t_0, r_\pm)
}
even though the first does not depend on $t_t=t_0$ the second does not depend on $\te_t=\te_0$.
We have two technical complications {to overcome}: there are different types of rays, each of which is differently parameterized.
{Indeed,} the signal source, {in the case of scattered rays,} could be {on} the same branch of the target or on the other.
{These possibilities yield 3} (in-falling, {scattered} on the same branch,  {scattered} on the opposite branch) $\times 2$ (ingoing or outgoing at the target)
$\times 2$ (counterclockwise or clockwise) light rays,  for a total of 12 types of light rays, each {of which is characterized by} an analytic expression {that can be examined} for possible intersections with the source worldline.
{However,} the parameterizations of the trajectories are complicated combinations of elliptic functions which are difficult to be solved analytically.
{Although we have analytic descriptions of rays and orbits, from the control problem on, we need to proceed numerically, which introduced some concerns about precision of computed values that need to be considered along the way.}
{For} these reasons, the search for the intersection is a bit convoluted.
In practice, {it is convenient to eliminate} the angle overcounting (encoded by the integer $n\in \Z$) by {solving} numerically two {of the} equations among
\eq{
 \cos(\te_l(r_s)) =  \cos(\te_s(r_s))
\qquad
 \sin(\te_l(r_s)) =  \sin(\te_s(r_s))
\qquad
 t_l(r_s) = t_s(r_s)
}
{We then verify that} the 3rd is also {satisfied. This approach eliminates} spurious solutions which correspond to the right cosine and the wrong sine, or vice versa.
%\EndNote

We fix the target (so that $(r_t, \te_t, t_t)$ become known parameters) {and} the source satellite (so that also $(\te_0, t_0, r_\pm)$
become known parameters). Thus, we are left with two unknowns {to solve for}, $(r_s, r_m)$ for {the scattered} light rays, $(r_s, K^2)$ for {the} in-falling rays. {The solution value} $r_m$ (or $K^2$) determines the ray, {while} $r_s$ determines the {point of} intersection {of} the ray and  the satellite worldline, {that is,} the source event. 

%\Note
{We} should also mention that, on {scattered} rays, it is convenient to compare {the values} $(\te_m, t_m)$ of the minimal approach event on the two branches of a ray passing through the source and the target so that one treats the two  {scattered} cases on {an} equal footing (and also because prolonging a ray to {another} branch accumulates numerical errors).

%\EndNote

If a solution is not found, one {needs to} explore the previous branch of the source orbit, as {happens in particular when the source satellite is near its} aphelion or perihelion.
Since the search is carried out on a type {by type} basis for rays, if a solution is not found on a {specific} type, one should search on {the} other types.

%\Note
One can speed up the search by noticing that $r_t>r_s$ and following clockwise the source along its orbit. We cycle among the rays of {the following} types: 
 in-falling-counterclockwise-outgoing, 
scattered on the same branch-counterclock\-wise-out\-going, 
scattered on other branch-counterclock\-wise-out\-going,
scattered on other branch-clockwise-out\-going, 
scattered on the same branch-clockwise-out\-going,
 in-falling, clockwise-outgoing.

Analogously, if  $r_s>r_t$, we cycle among the rays of {the following} types; 
in-falling-counter\-clock\-wise-ingoing, 
scattered on the same branch-counterclockwise-ingoing, 
scattered on other branch-counterclockwise-outgoing, 
scattered on other branch-clockwise-outgoing, 
scattered on the same branch-clockwise-ingoing, 
in-falling-clockwise-ingoing.

Two types {of rays} are not realized with all counterclockwise satellite orbits ({scattered} on other branch-clockwise-ingoing and  {scattered} on other branch-counter\-clock\-wise-ingoing).
Either way, we need to explore only six types of rays in each case instead of 12.

{Accordingly, we can start from the event on the source orbit, which has the same $t$ as the target, and then retrace backwards the orbit looking for solutions}. If we find a solution, we may continue to look for a better solution (one with a greater $r_m$). 
{Note that} changing the {orientation of the rotation} corresponds to the satellites passing behind the BH, and, consequently, to an event of the strong regime. 
{Strong field events are harder to {compute}, although they carry more information about the system, since the emission time is more sensitive to the characteristic of the system}. 

%\EndNote

The cases are undoubtedly complicated to {treat thoroughly}. However, {we} eventually, find several intersections corresponding to messages sent from satellites to the events, each {of which has its} {own} emission proper time $\tau_i$.

Depending on the situation, we can be satisfied with one message from each satellite (e.g., the one with a greater $r_m$)  or decide {to consider multiple messages} of the same satellites.
The choice depends essentially on how close a ray can go to the horizon and still be eventually detected.  
In an Earth setting, any ray hitting the surface at $r\sim 7\times 10^{7} \al$ cannot be later detected. 
In a BH setting (with no {accretion disk} at $r\sim \al$) we can {approach the horizon very closely} and be bent at will.
{Consideration of satellites around the Earth} is easier, but redundancy in the number of satellites {is required} to overcome the fact that the central mass can eclipse a pair of satellites. {This situation cannot arise} in a BH setting, since one can use strong lensing to avoid eclipses.

In any event, once we have determined the first two messages, {which can be referred to as} {\it first generation messages}, their emission is an event itself (on a satellite's wordlines), and one can look back for messages sent there from other satellites. We call these {\it second generation messages}, which, {are 4 in number}.
{It follows that, given each generation with $2^g$ emission events, we can go back to the previous generation's $2^{g+1}$ messages.  It is thus evident that this number}
increases exponentially with the number of generations.

There is also an equivalent formulation of the control problem based instead on {the} Hamiltonian formalism. 
We presented it in Appendix B, since it is a more general formulation of the control problem for a general Hamiltonian system, which {can be utilized} whenever we have a complete {integral} of the Hamilton-Jacobi equation.

\ 

\NewSection{Determining constellation geometry}

{We set up the positioning system in a region around the central mass by considering 3 freely falling satellites, {each one equipped with} an atomic clock.
They are able to exchange signals and mirror the signals they receive to the other satellites. 
We {assume that} the signals contain information about the clock reading at emission. 
Accordingly, at some point, a satellite receives a set of signals that have bounced back and forth among the satellites. 
In principle, each satellite receives an infinite sequence of proper times, organized in generations.} 

{We still need to discuss how we can extract, out of the infinite sequence of proper times $\tau_i$ available at a base event $p_0$ on a satellite, e.g., on satellite $\chi_0$, enough information to determine all orbital parameters of the satellite constellation, possibly together with  some gravitational parameters (as, e.g., $\al$ or perhaps some parameterization of the perturbations of the gravitational field to account for deviations from the Schwarzschild metric). We would also like to determine some additional orbital parameters (e.g., to model non-purely free falling satellites, {which take into account} the forces acting on the satellites due to solar wind or nuclear reactor thermal leaks as {occurred} with Pioneer's  acceleration anomaly),
 {and even} some parameters for the modification of the gravitational theory (as, for example, happens if we use Palatini $f(\calR)$-theories and {employ} Taylor coefficients of the function $f(\calR)$ expansion to parameterize the theory).}

One can regard these situations as discretizing the classical degrees of freedom for a generic gravitational field in a generic gravitational theory. If we keep the number of unknown parameters to be determined finite, there is no real difference between positioning, measuring the gravitational field, or discussing the observability of different gravitational theories.

{We} reason backwards.
The positioning of a user {can} easily {be} determined once the positions at the emission of satellites {are} known.

The user receives a signal from each satellite {which} contains information about the satellites' positions (in spacetime) at emission.
By knowing the position of the satellites in spacetime, the user {may conclude} {that he is on the future} light {conoid} at {the} emission position.
In dimension 2+1, {this} is a {2-dimensional} surface, and the user has three such surfaces. The intersection of 3 surfaces in a 3d manifold is a discrete set of points.
{If} we add redundancy (e.g.~increasing the number of satellites, restricting the domain of the chart, or remembering some information from previous positioning), the user can determine {his} position in spacetime, up to {an} {isometry}.

%\Note
{We} remark that the user does not {need to} use the clock readings it receives with the signal. {This means that} we do not need the clocks on satellites to be synchronized in any way {at this point}.
%\EndNote

Of course, for this scenario to work, each satellite should be able to determine its position in spacetime. 
This requires {that} the satellites {be able} to exchange signals.

\begin{figure}[htbp] %  figure placement: here, top, bottom, or page
   \centering
   \includegraphics[width=9cm]{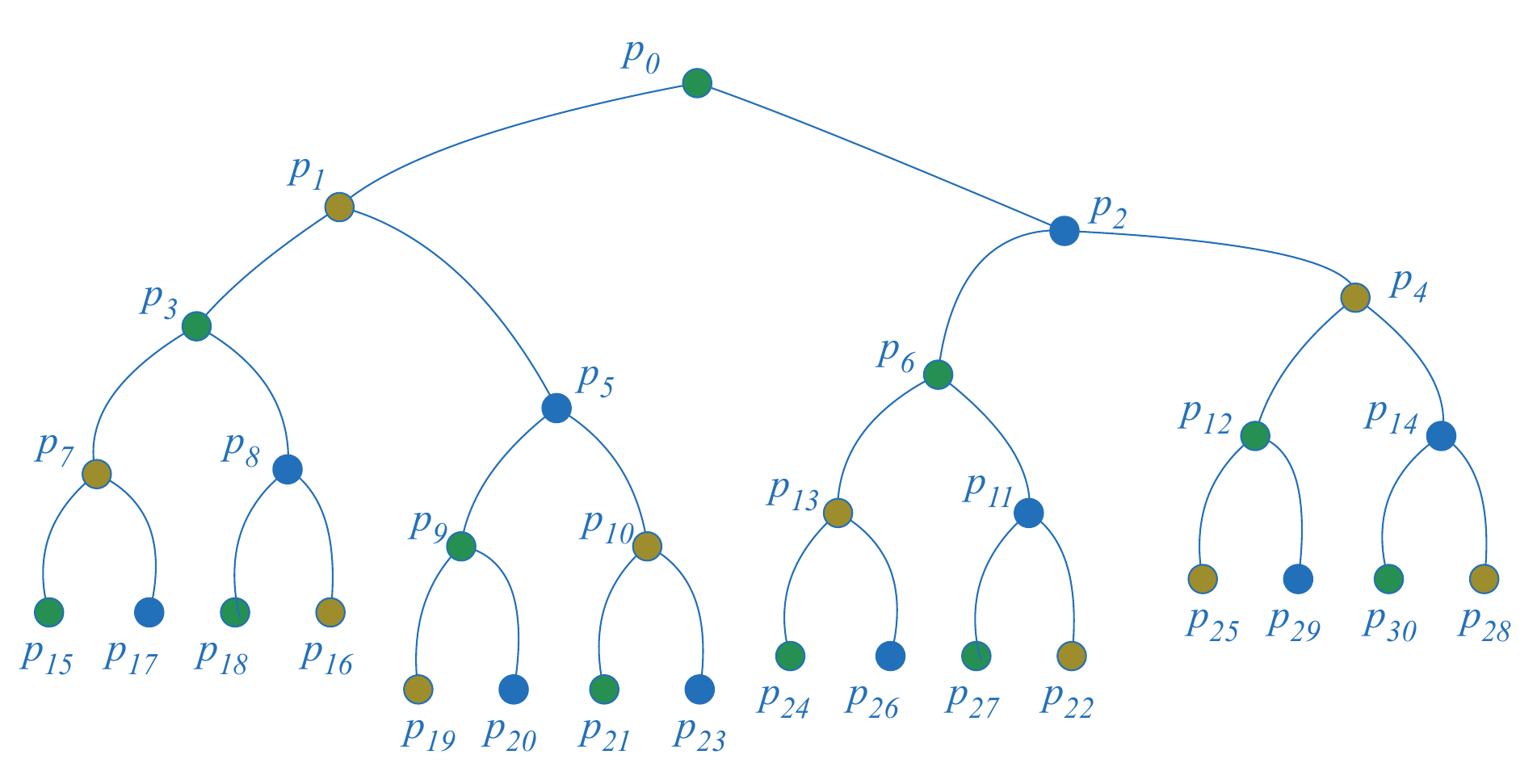}
   \caption{\small\it {The} first four generations of signals exchanged by three satellites  in dimension 3, considering a base event on $\chi_0$.
The event $p_A$ is on satellite $\chi_{A \> mod\> 3}$. Thus the signal from $p_{19}$ to $p_9$ is sent from satellite $\chi_1$ to $\chi_0$.}
 \end{figure}

{Suppose that} a satellite receives the sequence 
\eq{
(\tau_0; \tau_1, \tau_2; \tau_3,\tau_4,\tau_5, \tau_{6}; \tau_{7}, \dots, \tau_{14}; \tau_{15}\dots)
}
{which is the data from which it needs to determine the orbital parameters of all satellites.
On the other hand, the orbital parameters correspond to the degrees of freedom of the satellite constellation.}
{As we discussed in Section 2, the orbit of } each satellite {$\chi_i$} is described by {by five parameters $(t_i, \te_i, \tau_i, r^i_\pm)$, namely}
its initial  position {$(t_i, \te_i)$ on its current orbit branch} and clock reading at perihelion {$\tau_i$}, as well as the perihelion and aphelion of the orbit {$r_\pm$}. 

%\Note
If the  clock reading $\tau_e$ at emission  is given, then these orbital parameters completely determine the satellite's position at emission.
For example, given its orbital parameters, one can solve $\tau(r_e;r_\pm)=\tau_e$ to obtain $r_e$ at emission. 
Then $(r_e, \te(r_e; r_\pm), t(r_e; r_\pm))$ is the position at emission.
In other words, each satellite can compute its position by reading its clock, {\it once the orbital parameters are known}.

%\EndNote

{However}, because of isometries, we cannot obtain all {of the} orbital parameters {from the} signals. Since all signals are left invariant by rotations and time translations, we can fix the gauge by choosing $t_0=0$ and $\te_0=0$, i.e., giving all other positions with respect to the position of $\chi_0$.
We can also decide to make absolute clock readings irrelevant by setting {the} initial condition $\tau_i$ so that the proper time of $\chi_i$ 
at its first event $p_i$, {is} $\tau_i=0$. By doing this, we make any initial synchronizations irrelevant.

A particular sequence is entirely determined by the ($10= 15-3-2$) orbital parameters as well as the $r_0$ of the base event on $\chi_0$.
{Thus} we have 11 parameters as unknowns to parameterize the possible sequences of signals received at $p_0$.

%\Note
Given {the orbital parameters} $(r_0; \te_i, t_i, r^i_\pm)$, we have the base event $p_0=(r_0, \te(r_0; \te_0, r^0_\pm), t(r_0; t_0,  r^0_\pm))$.
As described in the previous Section, we find {the} emission events $p_1$ on $\chi_1$ and $p_2$ on $\chi_2$ which can send a light signal to $p_0$.
They are determined by $r_1$ and $r_2$ which completely determine $p_k=(r_k, \te(r_k; \te_k, r^k_\pm), t(r_k; t_k,  r^k_\pm))$, where 
$r_k= r_k(r_0; \te_i, t_i, r^i_\pm)$. {Thus we have determined} the first-generation signals. 

Given the events $p_1$ and $p_2$, we can look for signals $\chi_0\arr \chi_1$ and  $\chi_2\arr \chi_1$ to $p_1\in \chi_1$, which are emitted at $r_3$ and $r_5$, i.e.~at the events $p_3$ and $p_5$,
as well as   signals $\chi_0\arr \chi_2$ and  $\chi_1\arr \chi_2$ to $p_2\in \chi_2$, which are emitted at $r_6$ and $r_4$, i.e., at the events $p_6$ and $p_4$.
{Indeed, if} we write $r_A(r_0; \te_i, t_i, r^i_\pm)$ for each event, we can express the clock reading at emission as
$\tau_A(r_0; \te_i, t_i, r^i_\pm)=\tau_A (r_A(r_0; \te_i, t_i, r^i_\pm; \te_i, t_i, r^i_\pm)$. 
In other words, we can proceed backwards at will to compute {the} $\tau_A(r_0; \te_i, t_i, r^i_\pm)$ which are specific functions of the unknowns.
%\EndNote

In this way, we obtain an infinite {sequence} of equations which are invariant with respect to isometries and clock resettings, namely {we have}
\eqLabel{
\Cases{
 \tau_3( r_0, \te_i, t_i, r^i_\pm)= \tau_0-\tau_3 \cr
 \tau_4( r_0, \te_i, t_i, r^i_\pm)= \tau_1-\tau_4 \cr
 \tau_5( r_0, \te_i, t_i, r^i_\pm)= \tau_2-\tau_5 \cr
 \dots \cr
 \tau_A( r_0, \te_i, t_i, r^i_\pm)= \tau_{A \> mod \>  3} - \tau_A =:  \De \tau_A \cr
 \dots \cr
}
}{System}
Finally, we can imagine {that} each satellite broadcasts to the other satellites its clock reading as well as mirrors the signals which {it} is receiving at {the} same time from the other satellites.
From the point of view of a satellite, e.g., $\chi_0,$ which receives this transmission, then it is receiving precisely the sequence $\tau_A$ at the base event $p_0$. Since all {the} information $\tau_A$ is available to $\chi_0$ at $p_0$, {it} is able to write down the system (\ShowLabel{System}), solve it and determine the orbital parameters for all satellites as well as the base point $p_0$.

Each satellite, {following the same procedure simultaneously,} can compute its position (together with other information, e.g., multiple past positions of the other satellites and its own). This can be done at any moment by each satellite which can then broadcast to the user a signal encoding its position at emission. {In this way} the user, {upon} receiving {such messages from} each satellite, can determine its own position.

{One does} not need {an unlimited number of} equations to determine a finite number of unknown parameters. 
Thus, we can select some {$N$} initial equations, and keep the rest of the sequence for later use.

Unfortunately, the system (\ShowLabel{System}) cannot be easily solved; {thus, to proceed,} we turn it into a minimization problem for the function
\eqLabel{
\chi^2 ( r_0, \te_i, t_i, r^i_\pm) = \sum^{N+2}_{A=3} (\tau_A( r_0, \te_i, t_i, r^i_\pm)-\De \tau_A)^2
}{MinEq}
The solutions of the system (\ShowLabel{System}) correspond to absolute minima of the function $\chi^2$, {for} which one has $\chi^2=0$. 
{This result} may be helpful to exclude local minima which are not of interest here.

{A} solution $( r_0, \te_i, t_i, r^i_\pm)$ of (\ShowLabel{MinEq}) not only satisfies the first $N$ equations in system (\ShowLabel{System}),
but {also}, {whether} the initial assumptions we made (about the gravitational field, the motion of the satellites, and the gravitational theory we used) are {indeed} satisfied. {We note that, in addition,} all the other equations from the system (\ShowLabel{System}) have to be identically satisfied along the solution. 

If {this if is not the case,} the system {will} stop functioning as a positioning system to avoid broadcasting false information to the user.
Meanwhile, the satellites keep exchanging signals, computing orbital parameters, and checking {whether the} assumptions are eventually restored.
If they are, each satellite restarts broadcasting data to the user, which is able to {determine its} position again.
In this sense, the rPS is {\it self-locating} and {\it robust}.  

%\Note
Alternatively, satellites may test different hypotheses (e.g., transient non-gravitational forces {acting} on satellites {causing} them {to deviate} from timelike geodesics or transient deviations from {the} Schwarzschild metric {caused by}, for example, {an incident} gravitational wave). 
In this way, the system can autonomously try to model more general situations and self-determine extra parameters to model perturbations.
%\EndNote

Of course, this is only a concept toy model {which needs} to be made more precise. {However,} it shows how the rPS can be used as a gravitational detector and as a test for gravitational theories. 
{We} also remark that we have been {rather} vague about which parameters we fit because we {wish to preserve flexibility}.
If we are working around the Earth, we can {assume that} $\al\sim 0.008870\>m$.
However, if we dropped out of hyperspace near a black hole and we set up a rPS {there} to perform a physical experiment, we probably would like to {\it measure} $\al$ as well.

In that case, we {would} consider $\al$ to be an unknown {and}, if necessary, increase the number of generations {that} we use for determining the system parameters, and fit it with the orbital parameters. {We would proceed in a similar fashion}, if we {are expecting} perturbations of the gravitational field or want to test some non-standard gravitational theory.
The essential idea, however, is that we split {the} signals into two {sequences}: a first {sequence} used for fitting and {a second containing the remaining signals} {to be used} for checking {the} assumptions. The splitting time is called the {\it lookback time}, since only signals after the lookback time are used for the fitting.
What happens before the lookback time is irrelevant to the fitting. {Indeed}, we are assuming that deviations from assumptions are negligible {\it after} the lookback time.

\NewSection{Test the model: a BH case}

Using the techniques presented above, we are able to predict (exactly) the signals generated by a given constellation of satellites.
For example, we can consider {the} three satellites $\chi_i$.

Each satellite carries a proper clock, $\chi_0$ has orbital parameters $r_+=10\>\al$ and $r_-=19\>\al$. 
Its branch 0 starts at $(t_0=0\>\al c^{-1}, r_0=r_-= 10\>\al, \te_0=0)$ with   $\tau_0=1$.
One can compute that the $t$-coordinate period of $\chi_0$ is $2T_0=551.8586\>\al c^{-1}$, while its proper period is 
$2\tau_0=522.2874\>\al c^{-1}$, which shows the clock slows down due to the gravitational field. 
It {precesses} by $\de= 0.8736$ radians per orbit.

%\Note
{We} stress these quantities are calculated, not measured. Accordingly, we can provide a prediction {as} precise {as we wish}, e.g., with 120 digits.
Of course, a prediction {that is} too precise to be experimentally confirmed is {physically useless}.
{This} does not mean {that it does not have other uses.}
For example, {data are obtained in floating point with a  given precision, e.g., 40 digits and we have no idea on how reliable are the last digits.} 
However, it is simple to repeat the simulation at 120 digits and check how the digits {neighbouring} 40 behave. 
Moreover, {we note} that, in the case of the Earth, we can discuss situations ranging from one satellite orbit (about $10^{4}\> s$)
down to the accuracy of {an} atomic clock, that for {the} new optical clocks, can be expected to be less than $10^{-20}\> s$.
In other words, the situation we are considering naturally spans {a large} number of orders of magnitude.
Consequently, we need to guarantee at least the precision to describe {this situation}.
%\EndNote

The satellite $\chi_1$ has orbital parameters $r_+=28\>\al$ and $r_-=20\>\al$ with
$(t_1=10\>\al c^{-1},  \te_1=-\frac[7\pi/6], \tau_1=- \>\al c^{-1})$.
One can compute that the coordinate period of $\chi_1$ is $2T_1=1117.0454\>\al c^{-1}$, 
while its proper period is $2\tau_1=1081.5289\>\al c^{-1}$. It {precesses} by $\de= 0.4477$ radians per orbit.

The satellite $\chi_2$ has orbital parameters $r_+=40\>\al$ and $r_-=31\>\al$ with
$(t_2=-10\>\al c^{-1},  \te_2=\frac[3\pi/4], \tau_2=-3 \>\al c^{-1})$.
One can compute that the coordinate period of $\chi_2$ is $2T_2=1964.3794\>\al c^{-1}$, while its proper period is $2\tau_2=1922.4102\>\al c^{-1}$. It {precesses at} $\de=0.2886$ radians per orbit.

Then it is just a matter of computation (and {the selection of} one {light ray exchanged by two satellites} when multiple ones are available) to {complete} the table for signals described in Figure 2.
In this case, we have ({in units of} $\al c^{-1}$):
%\Note
{\small
\eqs{
\hbox{Generation 2}\quad
&\Delta \tau_3=   39.17850096193592366660462091592062373   \cr
&\Delta \tau_4=   80.11715717097871213365890494429573535    \cr
&\Delta \tau_5=   35.03574261053054630657492872958108159    \cr
&\Delta\tau_6=	 72.57341977793111680898418627081248044    \cr
\hbox{Generation 3}\quad
&\Delta \tau_7=  113.17356079751935436022513046580032987    \cr
&\Delta \tau_8=  35.40991007306522652038724051092347128  \cr
&\Delta \tau_9=  103.18980669701380595609210925548892818  \cr
&\Delta\tau_{10}=43.31148869147579252763214908406813583 	\cr
&\Delta \tau_{11}=65.43360770882868000737676330764741734    \cr
&\Delta \tau_{12}= 125.55127652440722347053216752374541211    \cr
&\Delta \tau_{13}=  80.13407626061348502470439015672952679    \cr
&\Delta\tau_{14}=112.24603172686866256539758944644350801	\cr
\hbox{Generation 4}\quad
&\Delta \tau_{15}=  159.84959178824144499228158441696501744   \cr
&\Delta \tau_{16}=   113.17357459679319449200407586632423227   \cr
&\Delta \tau_{17}=  93.10407540385622828546311154167323011   \cr
&\Delta\tau_{18}= 86.43058203591576981983712540155117890 	\cr
&\Delta \tau_{19}= 113.52586678836403966572945824855443803    \cr
&\Delta \tau_{20}=  143.91779026987465003621630041257875864    \cr
&\Delta \tau_{21}=  103.52001932858460948977835457701477422  \cr
&\Delta\tau_{22}=	 141.74429768374661292012973911416795685 \cr
&\Delta \tau_{23}=  76.85574698554547517515015464726496156  \cr
&\Delta \tau_{24}=    125.56902461398873586013930329151515120    \cr
&\Delta \tau_{25}=   185.53542787116689390175923961978786119   \cr
&\Delta\tau_{26}= 112.26226941709415618622735669007918004 	\cr
&\Delta \tau_{27}=   130.22536702210812661237475328349493992   \cr
&\Delta \tau_{28}=  136.84337647699999488633948073154505905    \cr
&\Delta \tau_{29}=  113.45578988671430465571931586496739842    \cr
&\Delta\tau_{30}= 172.61747721153048259619184107188535649  \cr
&\dots\cr
}

}

%\EndNote

{This is} the observed sequence, the shown digits {of which,} {are a} truncation of the actual computed values {to 35 decimal places}. 
{We remark} that these are observables of the {strictest} kind in GR; they correspond to {\it coincidences} in the theory,
of the sort {that} one can use to model observations at the most fundamental level.

{It remains} to show {that} we can actually solve for the orbital parameters, {and determine} {the precision and accuracy of the solution}. 
{This} is an issue precisely because we do not have much control over the expressions of the functions $\tau_A(r_0; \te_i, t_i, r^i_\pm)$
{or on} their differentiability. 
We shall {return to these questions for} the case of the Earth {considered} below. 
{We continue our analysis by determining} how the function $\chi^2$ behaves near a solution. 
We need to guarantee that the solution is an isolated minimum of $\chi^2$ and that the function is not too flat around it so that the minimizing algorithm can converge to the solution.

As an initial test, {perturb} the unknowns by a Gaussian noise {signal} ($\si=10^{-10}$) and check {to see} how the function $\chi^2$  {behaves.}
 We {observe} that, as expected, it responds by  $\De \chi^2\sim 10^{-17}$.
 If we {perturb} only the base event ($\si=10^{-10}$) on $\chi_0$, then we get $\De \chi^2\sim 10^{-21}$.
 If we {perturb} ($\si=10^{-10}$)  only satellite $\chi_0$,  $\De \chi^2\sim 10^{-16}$.
 If we {perturb} ($\si=10^{-10}$)  only satellite $\chi_1$,  $\De \chi^2\sim 10^{-17}$.
 If we {perturb} ($\si=10^{-10}$)  only satellite $\chi_2$,  $\De \chi^2\sim 10^{-17}$.
 If we {perturb} ($\si=10^{-10}$)  only one parameter of a satellite,  we get from 
 $\De \chi^2\sim 10^{-16}$ {to} $\De \chi^2\sim 10^{-23}$ depending on the {perturbation} and on the parameter we decide to {perturb}
 (besides, it obviously depends on the {unperturbed} configuration we start from).

\begin{figure}[htbp] %  figure placement: here, top, bottom, or page
   \centering
   \includegraphics[width=5cm]{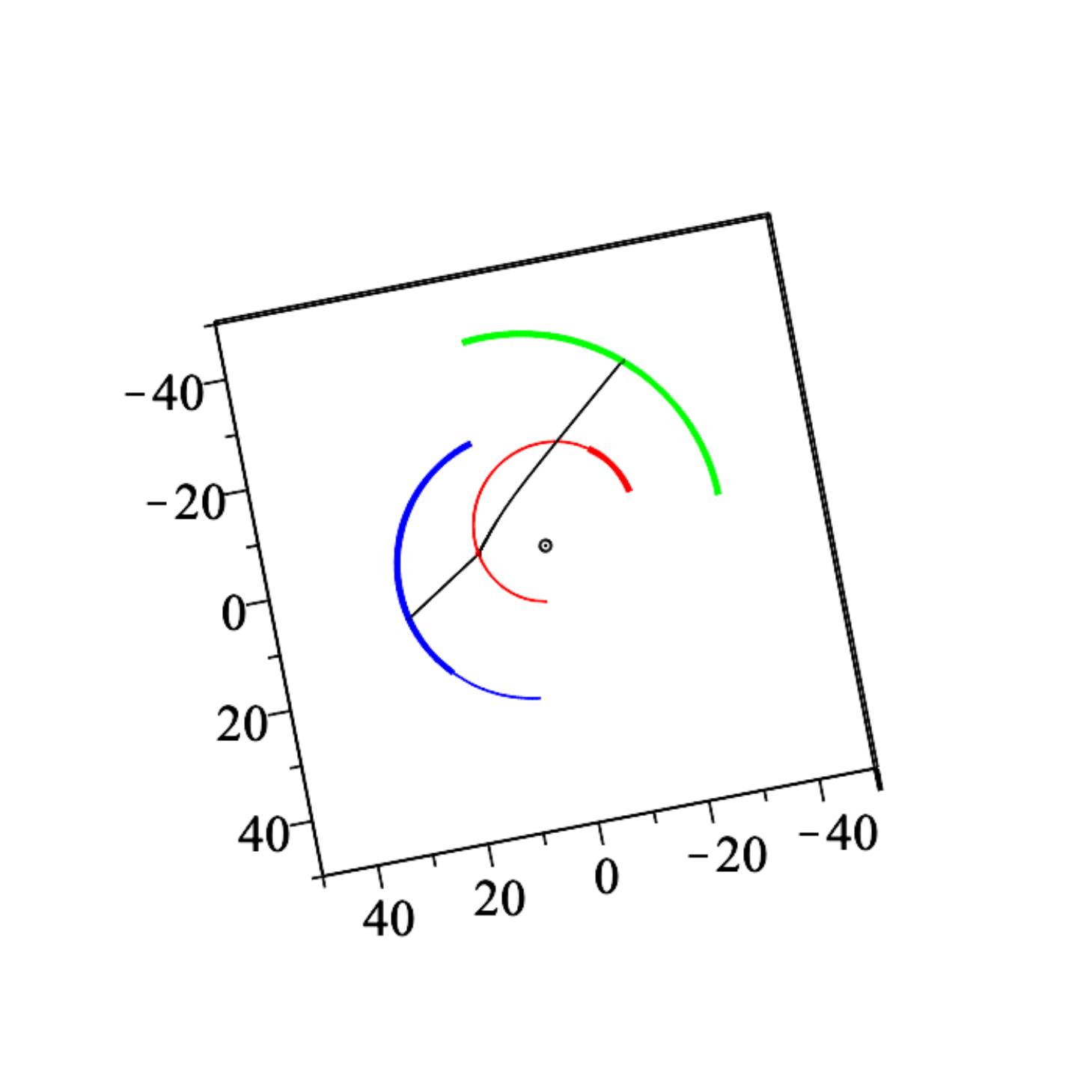}
   \qquad 
   \includegraphics[width=5cm]{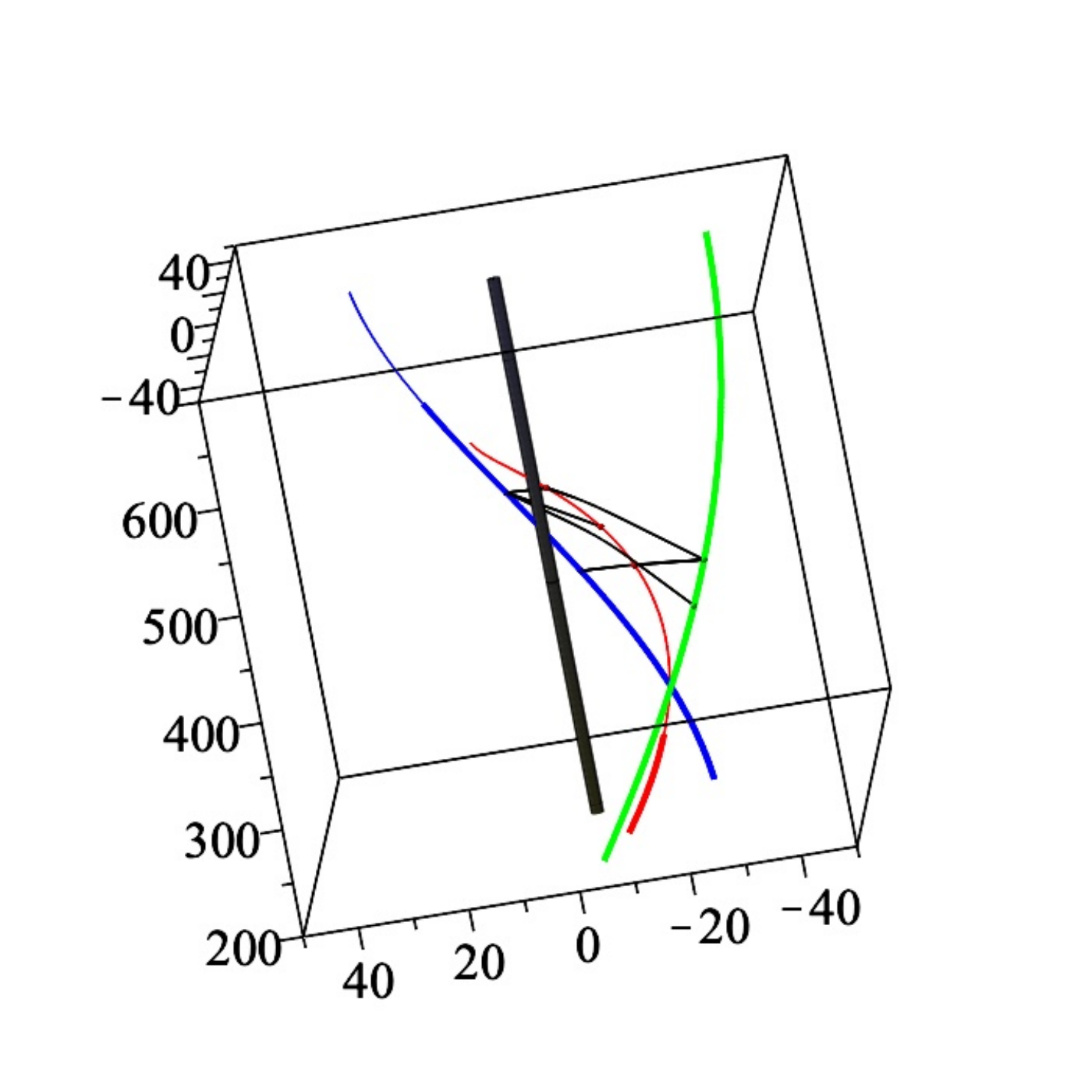}
   \caption{\small\it The signals of generation 1 on the left, of generation 1 and 2 on the right.}
\end{figure}

The response of the function $\chi^2$ to {a perturbation}, in this case, is {essentially} meaningless, because we are using homogeneous units 
that depend on the central mass.
However, in preparation for the Earth case, we need to know the precision required to  compute the function to determine the parameters.  
In {the} case {of a BH}, for example, we see that if we do not guarantee at least a precision to $10^{-20}$ in the value of $\chi^2$, we will eventually be unable to say {very} much. 
For example, in this case, we need a precision in measuring the signals $\tau_A$ at least of the order $10^{-10}$.

\NewSection{Test the model: an Earth case}

In the case of Earth, we model a situation similar to the actual GPS. {All results are} understood to be in SI units.

{We} consider 3 satellites $\chi_i$,
each {of which} carries a proper clock. $\chi_0$ has orbital parameters $r_-= 2.3289\cdot 10^{7}\>m$  and $r_+=2.4089 \cdot 10^{7}\>m$. 
{At} perihelion of branch 0 we have $\te_0=\frac[\pi/6]$, $t_0=-1500\> s$ and $\tau_0=-1500\>s$.

{The} coordinate period of $\chi_0$ is $2T_0=36284.7174 8955  \> s$, while its proper period (if we want to be able to {see} the difference) is $2\tau_0=36284.7174 7936  \> s$. {Comparison of these results shows that} the clock slows down due to the gravitational field. 
It {precesses} by $\de=3.5301\cdot 10^{-9}$ radians per orbit.

The satellite $\chi_1$ has orbital parameters $r_-=2.4289\cdot 10^{7}m$ and $r_+=2.5089\cdot 10^{7}m$.
For this {satellite} we have $\te_1=-\frac[\pi/6]$, $t_1=-1600\> s$ and $\tau_1=-1600\>s$.
The coordinate period of $\chi_1$ is $2T_1=38606.3642 2605\> s$, while its proper period is $2\tau_1= 38606.3642 1565\> s$. 
It {precesses} by $\de= 3.3871 \cdot 10^{-9}$ radians per orbit.

The satellite $\chi_2$ has orbital parameters $r_-=2.5289\cdot 10^{7}m$ and $r_+=2.6089\cdot 10^{7}m$.
{Further} we have $\te_2=0$, $t_2=-1400\> s$ and $\tau_2=-1400\>s$.
The coordinate period of $\chi_2$ is $2T_1=40975.5179 7465\> s$, while its proper period is $2\tau_2= 40975.5179 6404 \> s$. 
It {precesses} by $\de= 3.2551 \cdot 10^{-9}$ radians per orbit.
In this case, we have (all in seconds):
%\Note
{\small
\eqs{
\hbox{Generation 2}\quad
&\Delta \tau_3=   	0.177670545858865720368199359727407    \cr
&\Delta \tau_4=        0.006161739629485806051906548644238 \cr
&\Delta \tau_5=       0.057840923371459509637356702512431  \cr
&\Delta\tau_6=	      0.125990984860198315939016097030856\cr
\hbox{Generation 3}\quad
&\Delta \tau_7=        0.064002699645967673093551408614620  \cr
&\Delta \tau_8=        0.177670272908770755554022606444972  \cr
&\Delta \tau_9=        0.183831819371646273448205999862825  \cr
&\Delta\tau_{10}=     0.177670391080652676094777905921607 	\cr
&\Delta \tau_{11}=    0.125990791303975115694999501565061   \cr
&\Delta \tau_{12}=    0.183832280120454500258023229898599 \cr
&\Delta \tau_{13}=    0.125990875102861017455161133288578 \cr
&\Delta\tau_{14}= 	0.064002666904713921376083935866949 \cr
\hbox{Generation 4}\quad
&\Delta \tau_{15}=    0.241673189747825261701984791678735  \cr
&\Delta \tau_{16}=    0.183831659226008694553003310344026  \cr
&\Delta \tau_{17}=    0.183831536956154698351275644437245 \cr
&\Delta\tau_{18}= 	0.355340782158789752780184396885066 \cr
&\Delta \tau_{19}=    0.183832125100949090748893101070209 \cr
&\Delta \tau_{20}=    0.121843663566318426499010983039103 \cr
&\Delta \tau_{21}=    0.303660984816986754999415382305216  \cr
&\Delta\tau_{22}=	0.132152610754713273885450011424911 \cr
&\Delta \tau_{23}=    0.235511427015118169710635646895444 \cr
&\Delta \tau_{24}=    0.303661311202605558019309110571608    \cr
&\Delta \tau_{25}=    0.070164447082984281655998663824688   \cr
&\Delta\tau_{26}= 	0.183831878295795158020429178386318  \cr
&\Delta \tau_{27}=    0.251981582606611857425290985243845    \cr
&\Delta \tau_{28}=    0.183832119974415537678739505974830  \cr
&\Delta \tau_{29}=    0.183831997704255089293768828935526 \cr
&\Delta\tau_{30}=     0.189993553438716616270862526046800\cr
&\dots\cr
}
%\EndNote
}
{We} stress {that these are the results of} {\it one} simulation for more or less generic parameters. 
To solve {the} positioning {problem} we need to {sample parameter space} {with a significant number of simulations}. 
{The results are discussed} below.

{We begin by repeating the perturbation testing.}
 {We perturb the unknows} by a Gaussian noise {signal} ($\si=10^{-10}\> m$) and {examine} how the function $\chi^2$ responds 
 by  $\De \chi^2\sim 10^{-20} \> s^2$.
 
 If we {perturb} only the base event ($\si=10^{-10}$) on $\chi_0$, we get $\De \chi^2\sim 10^{-33} \> s^2$.
 If we {perturb} ($\si=10^{-10}$)  only satellite $\chi_0$,  $\De \chi^2\sim 10^{-22} \> s^2$.
 If we {perturb} ($\si=10^{-10}$)  only satellite $\chi_1$,  $\De \chi^2\sim 10^{-22} \> s^2$.
 If we {perturb} ($\si=10^{-10}$)  only satellite $\chi_2$,  $\De \chi^2\sim 10^{-24} \> s^2$.
 
 If we {perturb} ($\si=10^{-10}$)  only one parameter of a satellite,  we get from 
 $\De \chi^2\sim 10^{-21}\>s^2$ {to} $\De \chi^2\sim 10^{-37}\>s^2$ depending on the {perturbation} and on the parameter we decide to {perturb}
and the configuration we start from.

%\Note
{This needs extra care since $\De\chi^2$ is quite close to the precision at which we are computing everything.
So we went back and recomputed everything with a precision $10^{-80}$ instead of $10^{-40}$. {We were able to confirm} that these figures are really there.}

We see that, with a clock sensitive to $10^{-20}\>s$, we can measure $\chi^2$ with a precision {near to} $10^{-40}$.
{This} corresponds to {a} change of parameters of {approximately} $10^{-20}$.
% \EndNote
  
{Indeed}, this is the best we can hope {for}, not necessarily what we {can} manage to attain. {However, we note that,} regarding the positions, we are talking about  $1\> mm$ for a satellite $2\cdot 10^7\> m$ away!

   \begin{figure}[htbp] %  figure placement: here, top, bottom, or page
      \centering
      \includegraphics[width=9cm]{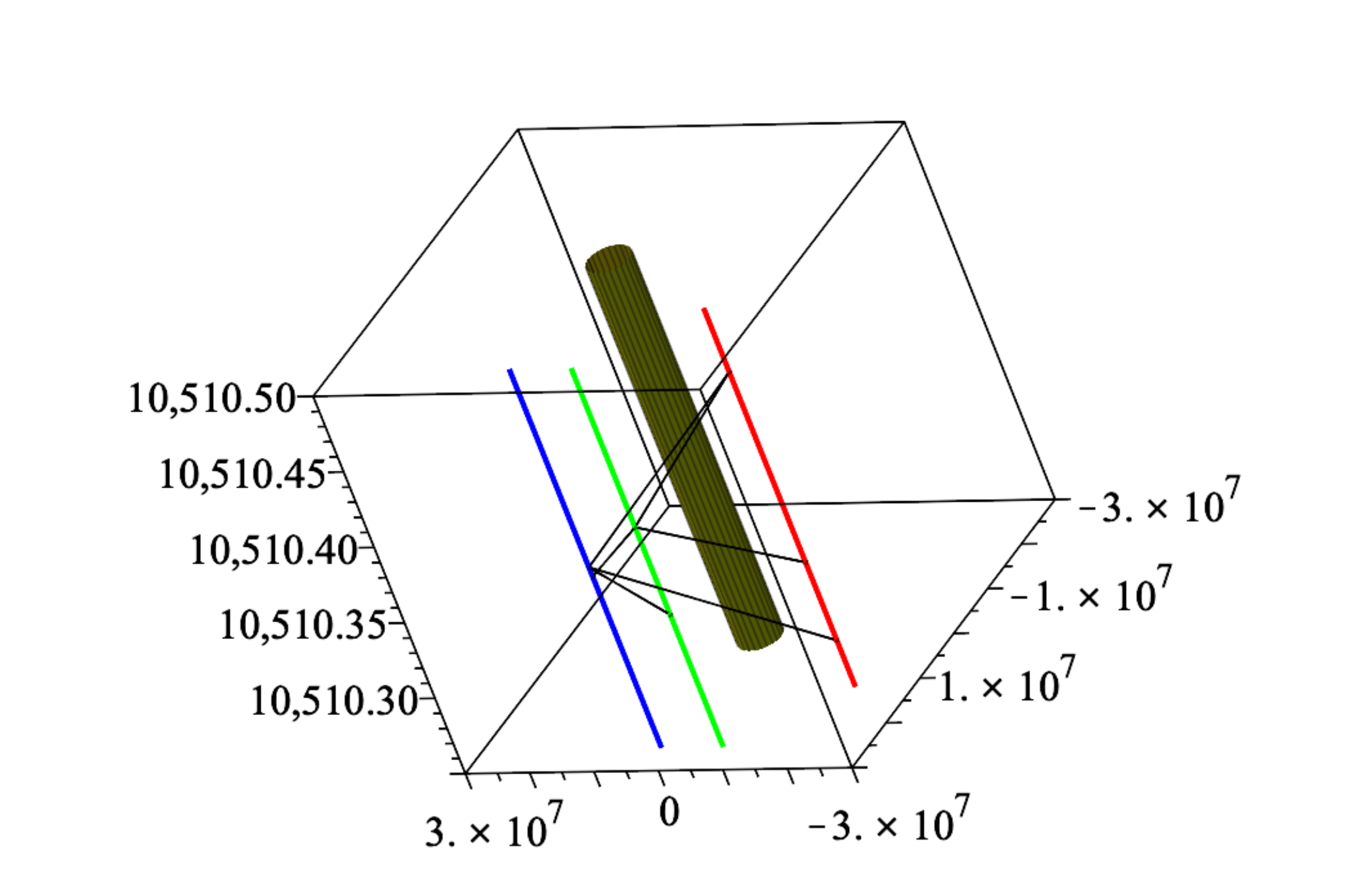} 
      \caption{\small\it The signals of $2^{nd}$  and $3^{rd}$  generations}
   \end{figure}

{We summarize by remarking that we have been able to reduce} the whole issue of orbital parameters determination to a minimization.
It turns out {that} we have several very good algorithms to minimize a function $\chi^2$ even when we do not know much {regarding} {its} differentiability.
After all, that is what machine learning is all about.

\NewSection{Conclusions and perspectives}

{In summary, we {show} that in dimension 2+1 we can cope with problems we did not have in 1+1.
We are able to analytically describe the exact bounded orbits to model the satellites 
and light rays to model signals exchanged between them.
We are able to solve the control problem in quite a general way, by determining the signal emitted from a satellite to a target event, possibly on another satellite.
Finally, the simulation phase is robust enough to actually minimize the function, {from} which {one} can determine {the} orbital parameters.}

We {would like} to provide some preliminary evidence that, in principle, one can solve the minimizing problem for the function $\chi^2( r_0, \te_i, t_i, r^i_\pm)$.
There are {many} algorithms {available} to do {this computation;} {they} mainly differ in efficiency and the regularity required for the function.
In particular, there are several algorithms which rely only on function evaluations. 
{We} remark that we are using a parameter space which is 11 dimensional. {Since} we plan to perform {the} minimization continuously, {we would like} to use the previous solution found as a guess for the solution of the following iteration.

We expect that {the} performance {can be improved} by tailoring the algorithm to the specific situation we are considering.
However, at this stage, we are not concerned with efficiency. 
We took the standard minimizer of Maple (in the {\tt Optimization} package) with {a} little juggling on its configuration. 
Then we {model} rPS in the Earth situation,
{we set} {the mathematical precision} to $10^{-80}$, 
{and} the orbital parameters as described above.
{With this setup} we simulate signals so that we know a solution quite precisely.

{We proceed by perturbing} the solution  ($\si=10^{-5}$) to simulate a previous solution near it.
We want to show that the minimizer can start from this initial {perturbation} and converge to the known solution. 
{We ask} for a precision of $10^{-20}$ (for the function $\chi^2$), which corresponds to a precision of about $10^{-10}$ {for the} parameters.
The Maple minimizer returned, as expected, to the solution, which is quite close to the known solution, namely 
we {observe that} each parameter {starts} from the {perturbed} value and {ends} near the known solution. 
{The results of the calculation are given} in the following table
of the absolute accuracy of {perturbed values and the final minimum found with respect to the calculated solution}. 
\medskip

\hbox to \hsize {\hfill\vbox{\offinterlineskip \tabskip=0pt \halign{
\vrule height2.75ex depth1.25ex width 0.6pt #\tabskip=1em &
\hfil 0.#\hfil &\vrule # & \qquad$0.#\,\pi$\hfil &\vrule # &
\hfil 0.#\hfil &#\vrule width 0.6pt \tabskip=0pt\cr
\noalign{\hrule height 0.6pt}
&\omit       &&\omit $\De$perturbed && \omit $\De$minimum & \cr
\noalign{\hrule}
&\omit $r_0$ 	&&\omit 	$0.57987\cdot 10^{-5}$  	 &&\omit 	$1.0\cdot 10^{-10}$	 &\cr
&\omit $r^0_-$ 	&&\omit 	$1.0037\cdot 10^{-5} $ 	 &&\omit 	$1.0\cdot 10^{-10}$	 &\cr
&\omit $r^0_+$ &&\omit 	$1.8841\cdot 10^{-5}$	 &&\omit 	$1.0\cdot 10^{-10}$	 &\cr
&\omit $\te_1$ 	&&\omit 	$2.8649\cdot 10^{-7}$ 	 &&\omit	$-0.4096\cdot 10^{-10}$ &\cr
&\omit $t_1$ 	&&\omit 	$0.23341\cdot 10^{-5}$ 		 &&\omit 	$1.0\cdot 10^{-10}$	 &\cr
&\omit $r^1_-$ 	&&\omit 	$1.662\cdot 10^{-5}$ 		 &&\omit 	$1.0\cdot 10^{-10}$	 &\cr
&\omit $r^1_+$ &&\omit 	$-1.2022\cdot 10^{-5}$ 		 &&\omit 	$1.0\cdot 10^{-10}$	 &\cr
&\omit $\te_2$ 	&&\omit 	$-2.3878\cdot 10^{-5}$ 		 &&\omit 	$1.0\cdot 10^{-10}$	 &\cr
&\omit $t_2$ 	&&\omit 	$0.66986\cdot 10^{-5}$  		 &&\omit 	$1.0\cdot 10^{-10}$	 &\cr
&\omit $r^2_-$ 	&&\omit 	$0.93391 \cdot 10^{-5}$ 		 &&\omit 	$1.0\cdot 10^{-10}$	 &\cr
&\omit $r^2_+$ &&\omit 	$0.58673\cdot 10^{-5}$ 	 &&\omit 	$1.0\cdot 10^{-10}$	 &\cr
\noalign{\hrule height 0.6pt}
} }\hfill} 
\medskip
%\Note
In order to compute the minimization the algorithm has to do about 100 evaluations {of the function $\chi^2$}, each of which {requires that} a complete simulation  {is} made.
Currently, each simulation takes about $60\>s$, namely, about $2\>s$ per signal.
{By the use of} more performance-oriented programming languages ({\tt Swift} and {\tt C++}, using GSL libraries) we obtained better performance  {by} about a factor {of} 100, {albeit} at the price of having less control on precision and accuracy and of doing {the computations} numerically.
Here we preferred to go for accuracy and precision, neglecting performance, since we are discussing viability {\it in principle}.
Maple's simulation can be found at \cite{link}.
%\EndNote

{We stress that} even without juggling much with its configuration, the algorithm converges (though quite slowly) to the solution. 
The main point here is that the minimizer requires about 100 evaluations to run. Thus the main challenge was to obtain
a robust simulation procedure, which allows us to evaluate the function $\chi^2$ for all initial conditions in a neighbourhood of the solution. {The results} indicate that the function $\chi^2$ is well-defined and continuous almost everywhere near a solution.

{We} are able to argue that the precision of the clocks, {which} sets a limit to our ability to measure $\chi^2$,
{implies}  a limit in the precision {with which} we can determine, {\it in principle}, the orbital parameters.
In the Earth configuration, we can reasonably imagine an order of magnitude in the range of {$10^{-3}\>m$.}

%\Note
{Under} the best conditions, we can {envisage} an optical clock with a precision of {approximately} $10^{-20}\> s$ which allows us the measure $\chi^2$
to order $10^{-40}\> s^2$. {This precision leads} to orbital parameters {of order} $10^{-20}$.
For the error {in} position, we have $(\De r)^2= \De r^2 + r^2\De\te^2$ {which implies} an error of {approximately}  $\De r\sim 10^{-13}\>m$.
{These results are adequate for a} completely theoretical estimate, {since} currently  {this scenario is not yet} possible.
{It can be considered as} an approximate maximal precision estimate.
%\EndNote

{In this regard we} remark that many other sources of error affect the final result, first of all, atmospheric turbulence.
{This} is why we {have} designed the process of determining orbital parameters {in order} that it is separated from {the determination of} user positioning, namely, it happens entirely out of the atmosphere.

Further investigation needs to be devoted to the domain in which our procedure is working, as well as for a more realistic evaluation of errors, for which one needs details about {the} satellites {that are} unavailable at the moment.

{At this point of our research} we have learned that a relativistic theory provides us with the tools to measure the motion of objects around a source so that we can bootstrap laws of dynamics without relying on Newtonian physics or special relativity. 
{The information obtained may be used} to measure the gravitational field (at least in a discretized version of it) by fitting perturbations of Schwarzschild,
and to discuss the observability of modified gravitational theories in terms of explicitly observable quantities.

We {are confident} that the whole derivation can be carried out for other solutions of Einstein equations, e.g., {the} Kerr spacetime,
provided {the solutions possess} enough first integrals to determine {the} geodesic trajectories in terms of Weierstrass equations, which corresponds to having a complete {integral} of the Hamilton-Jacobi equation.

\NewAppendix{A}{drawing the light conoid} 

Using equations of light rays (\ShowLabel{LightRaysWeierstrass}), we can fix the vertex event $p_0=(r_0, \te_0, t_0)$ and parameterize the past {light conoid} by $(r, r_m)$ (or $(r, K^2)$ depending on the surface region). 

\begin{figure}[htbp] %  figure placement: here, top, bottom, or page
   \centering
   a) \includegraphics[width=5.3cm]{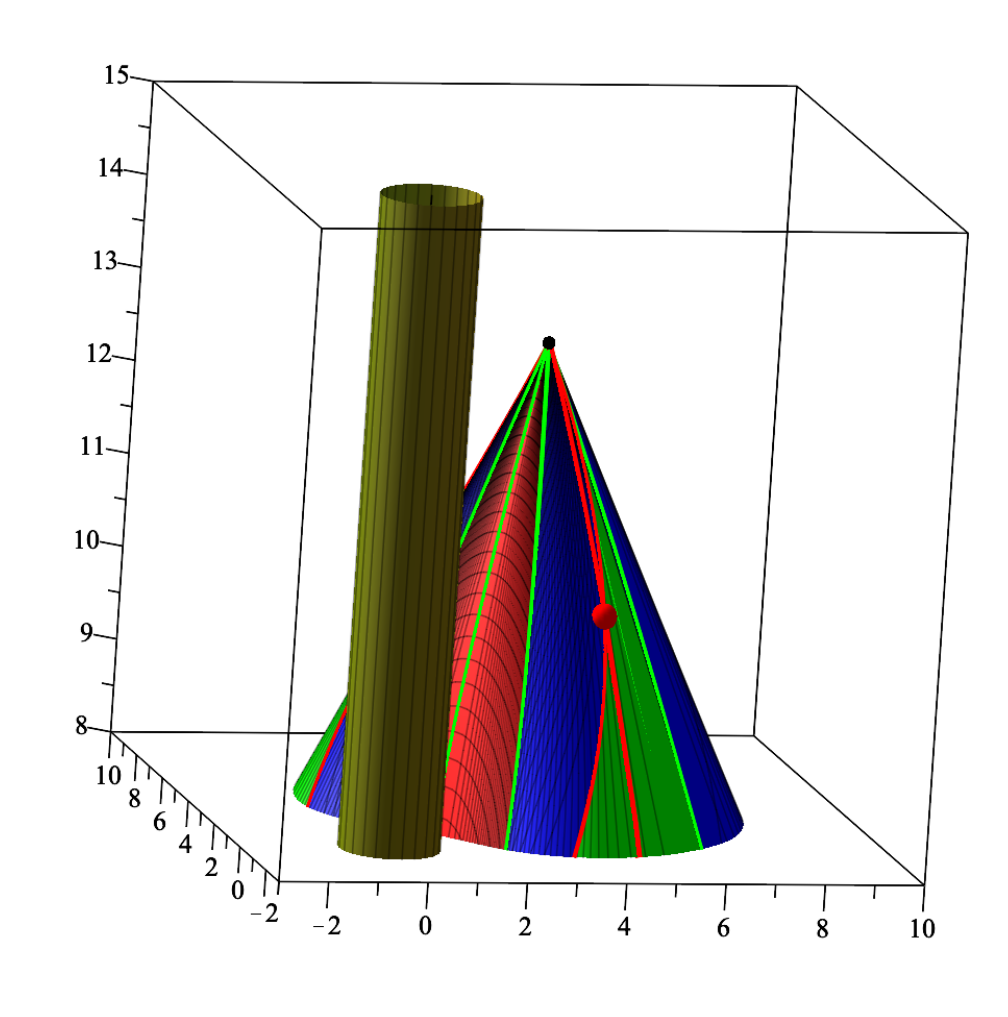}\\ 
   b) \includegraphics[width=5.3cm]{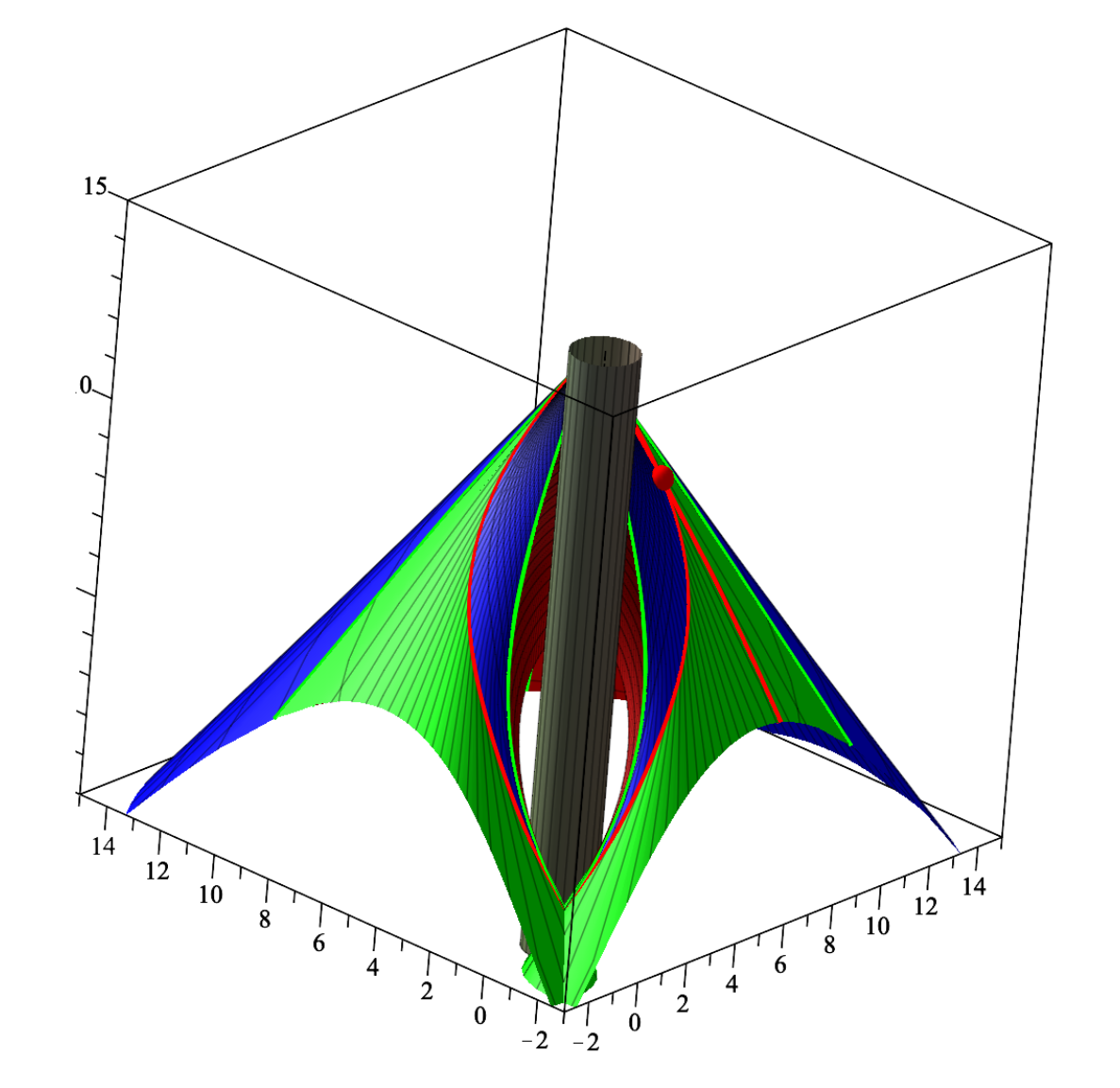}\\ 
   c) \includegraphics[width=5.3cm]{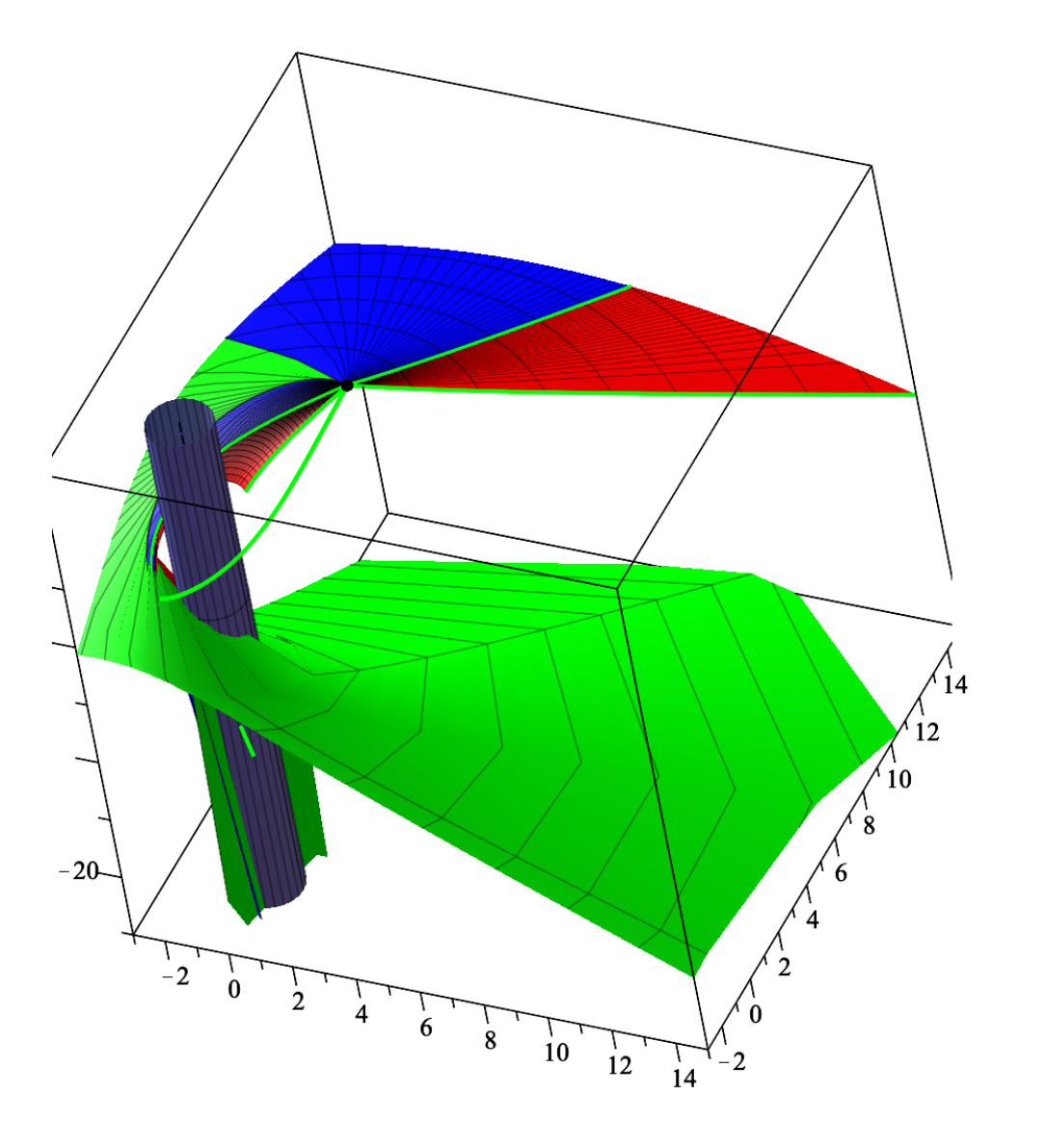} 
\caption{\small\it The three figures show the light {conoid}, precisely the past {conoid} of the event $p_0=(r=5, \te=\frac[\pi/4], t=13)$.
The red regions are the in-falling rays, the blue regions the {scattered} rays {in the} branch containing $p_0$, the green parts {are} the opposite branches
of {the scattered} rays.
Lengths are measured in units $\al$, times so that $c=1$. The central cylinder is the BH horizon.
\goodbreak
a) {shows} the interval $t\in [8,15]$. Here the surface is a cone with a vertex in $p_0$ deformed by the BH bending of light and synchronization.
The red line shows a ray through $p_0$, which reaches the minimal spatial distance at the red dot and then goes out on the other branch.
\goodbreak
b) {shows} the  interval $t\in [-5,15]$. We see {that the generators of the light conoid self-intersect} after going around the BH.
\goodbreak
c) {shows} the  interval $t\in [-25,15]$ {with} two sheet of the light {conoid.}}
\label{figConoidExs}
\end{figure}

{As an} example, we can consider {scattered}, ingoing, counterclockwise, light rays through $p_0$ which are parameterized as
\eq{
r = r
\qquad\qquad
\te = \te(r, r_m) 
\qquad\qquad
t = t(r, r_m)
}
where the functions $\te(r, r_m)$ and $t(r, r_m)$ are given by the solution in (\ShowLabel{SolutionScatteringRays}) (in this case with $e=-1$ and $\si=1$).
Accordingly, we can draw the corresponding region of the light {conoid}, for clockwise rays, for counterclockwise and clockwise outgoing rays (see Fig. \ref{figConoidExs}).
When we consider outgoing rays, we can trace back on the surface the curve on which they reach their minimal $r=r_m$ from the BH,
and draw, for each outgoing ray, its ingoing branch, which fills another region on the light {conoid}.
Finally, we can consider in-falling (incoming and outgoing) rays, which give us two more regions of the light  {conoid} using the solutions (\ShowLabel{SolutionInfallingRays}).
{We observe} that since we have analytic solutions, we are drawing the exact analytic light {conoid} with no approximations.

It is interesting to draw the {past} light {conoid} up to some time $\de t$ {antecedent} to its vertex $p_0$.
In this case, for small enough $\de t$, the surface appears to be a conoid, as expected.
If we then increase $\de t$, multisheets appear. 
{By} definition of the {light conoid}, each {of its} points is connected to the vertex by a light ray. {Thus,} if we consider a particle $P$ with a worldline which intersects the light {conoid} {with vertex $p_0$}  more than  {once}, we have a light ray starting at each intersection and reaching {$p_0$}. 
{This situation} shows {that} the vertex sees multiple images of the particle $P$ emitted at different times although received at the same time.

\NewAppendix{B}{Hamiltonian control problem}

The control problem {can be extended} to a general Hamiltonian system. 
The problem {consists of the determination of}  a solution {through given initial and final positions}  by finding the initial (and final) momenta.

Hamiltonian systems are defined on {a symplectic manifold, e.g.~} the cotangent bundle $T^\ast Q$ of the configuration space $Q$ which is parameterized by {the} position {coordinates $q$}. The dynamics is described by a Hamiltonian $H(q, p)$ which can be determined by {the} Legendre transform of the Lagrangian. 
In the case of the Lagrangian (\ShowLabel{Lag}) {the} momenta are defined {by}
\eq{
p_r= -\Frac[ \dot r/\sqrt{A(   c^2  A^2  -  \dot r^2  - A r^2  \dot\te^2 )  }]
\qquad
p_\te = -\Frac[ A r^2\dot\te /\sqrt{A(   c^2  A^2  -  \dot r^2  - A r^2  \dot\te^2 )  }]
} 
and the Hamiltonian {by}
\eq{
H(r, \te, p_r, p_\te)= \Frac[ c\sqrt{A} / r]  \sqrt{ r^2 +Ar^2 p_r^2+p_\theta^2 }
}

Since the Hamiltonian is independent of $t$ and $\te$ is cyclic, one can look for a complete {integral} of the Hamilton-Jacobi {(HJ)} equation in the form
\eq{
S(t, r, \te; E, J)= -{c^2}Et+J\te + W(r; E, J)
}
{It follows that} the {HJ} equation {reduces to}
\eq{
\Frac[dW/dr] = \pm\Frac[\sqrt{ ({c^2}E^2-A)r^2-J^2 A}/r A]
}
%\notaL{To get this form $E$ has to be replaced by $c^2E$ in $S$ above.}

{Thus the} solution $W$ (and thence a complete integral $S$) {is given by}
\eq{
W(r; E, J)=\pm \int \Frac[\sqrt{ ({c^2}E^2-A)r^2-J^2 A}/r A] dr
}
{where the value of the integration constant is immaterial.}

The complete integral $S(t, r, \te; E, J)$ is a generating function for the canonical flow of transformations
\eq{
\Phi_S : (t, r, \te, p_r, p_\te)\mapsto (t, E, J, P_E, P_J)
}
where the conjugate momenta $(P_E, P_J)$ are also first integrals.

%\Note
{We note} that, for any Hamiltonian system, one has a complete integral of {the} Hamilton-Jacobi equations.
{Indeed}, the evolution of the system is always a canonical flow, {since} the complete integral is nothing but the generating function
of the evolution flow. 
In fact, one can consider the action $A=\int_{t_0}^{t_1} L\circ \si dt$ computed along solutions $\si$ with a fixed initial and final positions, 
write it as a function of initial and final positions $S(t_0,q_0, t_1, q_1)$, which is called the {{\it Hamilton's principal function}}, and show directly {that} it satisfies {the} Hamilton-Jacobi equation. {We note that the distance-function defined in Section 2 is the square of Hamilton's principal function \cite{Synge2}}.

{This proves} existence. {However,} to compute Hamilton's {principal} function {in} this way one needs first to know the general solution of the system. However, if {such a solution is unavailable}, we still may be able to {find a complete integral}, e.g., by separation of variables, 
and {then} show {that it satisfies the HJ equation}. {As mentioned above} the complete integral generates the evolution flow, hence the general solution of the system. 
{Thus}, given a complete integral of the {HJ} equation, we can construct the  Hamilton's {principal} function, which, in fact, is a kind of classical propagator, {since} it solves the control problem of finding solutions that go from $(t_0, q_0)$ {to} $(t_1, q_1)$.
%\EndNote

We can compose two such flows
\eq{
\Phi_F: (t, r, \te, p_r, p_\te)\mapsto (t, E, J, P_E, P_J) \mapsto  (t_0, r_0, \te_0, p^0_r, p^0_\te)
}
associated with the generating function
\eq{
F(t, r, \te; t_0, r_0, \te_0)= S(t, r, \te; E, J) - S(t_0, r_0, \te_0; E, J)=: S- S_0
}
{This} is called the {\it evolution generator}.

%\Note
Since we know that $(P_E, P_J)$ are conserved along solutions one has
\eqLabel{
\Cases{
  \Frac[\del F/ \del E] =  \Frac[\del S/ \del E] -  \Frac[\del S_0/ \del E] = P_E-P_E =0\cr
   \Frac[\del F/ \del J] =  \Frac[\del S/ \del J] -  \Frac[\del S_0/ \del J] = P_J-P_J =0\cr
}
}{FEq}

Then, in principle, one can solve this system for $E=E(t, r, \te; t_0, r_0, \te_0)$ and $J=J(t, r, \te; t_0, r_0, \te_0)$, {and} replace them back into $S-S_0$ to obtain $F$ as a function of 
$(t, r, \te; t_0, r_0, \te_0)$ only.

The evolution generator $F(t, r, \te; t_0, r_0, \te_0)$ obtained in this way represents a congruence of solutions (i.e.~geodesics) {which start} at $t_0$ 
from the point $(r_0, \te_0)$
(i.e.~from the event $( t_0, r_0, \te_0)$) {and go} to the event $(t, r, \te)$ {which corresponds} to the first integrals $\(E(t, r, \te; t_0, r_0, \te_0), J(t, r, \te; t_0, r_0, \te_0)\)$.
In other words, it solves the typical control problem in which we control the initial and final position and determine initial and final momenta (or equivalently, the first integrals $(E, J)$)
needed to go from {the} initial to the final position (see \cite{Benenti}).

 Finally, {we} remark that if $E$ is negative, we have timelike geodesics, {while} if $(E, J)$ goes to infinity {and the ratio}
 \eq{
 K:= \Frac[J / E] = \Frac[r^2 \dot \theta / A] = - \Frac[ c r p_\te/ \sqrt{A(p_r^2 r^2 A+p_\te^2+r^2)}]
 }
 {remains} finite, we have  {\it light rays}.

%\EndNote

Hence, we have a {procedure} to obtain an explicit description of light rays. 
We consider the equations (\ShowLabel{FEq}), we replace $J= K \ep^{-1}$ and $E=\ep^{-1}$ and expand in series of $\ep$. 
Since we want solutions {where} $E\arr -\infty$, (i.e.~$\ep \arr 0$), one necessarily needs the zero-order term in the series to vanish, otherwise, $\ep=0$ would not be a solution.
{This process yields} the equations
\eqLabel{
\Cases{
  \Frac[\del F/ \del E] (t, r, \te; t_0, r_0, \te_0; K, \ep=0) =  0\cr
   \Frac[\del F/ \del J] (t, r, \te; t_0, r_0, \te_0; K, \ep=0) =  0\cr
}
}{LR}
{the solutions of which:} $t=t(r;  t_0, r_0, \te_0, K)$, $\te= \te(r;  t_0, r_0, \te_0, K)$, {determine} a light ray through $(t_0, r_0, \te_0)$.
Different values of $K$ parameterize different ``types'' of light rays (since the value of $K$ is constant along a single light ray).

{Thus, given} two events $( t_0, r_0, \te_0)$ and $(t, r, \te)$, it is easy to {verify} whether there is a light ray {joining} them:  {one checks whether} there is a value of $K$ such that equations (\ShowLabel{LR}) are satisfied, i.e., if they {have a solution} as equations for $K$.
 
The typical problem we have to solve for computing signals is {given} a fixed event $P=(t_0, r_0, \te_0)$ and a timelike curve $\ga: \R\arr M: s\mapsto (t(s), r(s), \te(s))$,  find a point on the trajectory of $\ga$ such that it is connected to $P$ by a light ray.
Then we simply write down equations (\ShowLabel{LR}) as
\eqLabel{
\Cases{
  \Frac[\del F/ \del E] (t(s), r(s), \te(s); t_0, r_0, \te_0; K, \ep=0) =  0\cr
   \Frac[\del F/ \del J] (t(s), r(s), \te(s); t_0, r_0, \te_0; K, \ep=0) =  0\cr
}
}{SignalEq}
and solve them for $(s, K)$. The value of $s$ gives the event $Q=(t(s), r(s), \te(s))$ on the trajectory of $\ga$, {while} the value of $K$ selects the light ray through $P$ and $Q$.

{\small

\Acknowledgements
This article is based upon work from COST Action (CA15117 CANTATA), supported by COST (European Cooperation in Science and Technology).

We also acknowledge the contribution of INFN (Iniziativa Specifica QGSKY and Iniziativa Specifica Euclid), the local research project {\it  Metodi Geometrici in Fisica Matematica e Applicazioni (2022)} of Dipartimento di Matematica of University of Torino (Italy). This paper is also supported by INdAM-GNFM.

L. Fatibene would like to acknowledge the hospitality of the Department of Applied Mathematics, University of Waterloo, where part of this research was done.
This work was  supported in part by a Discovery Grant from the Natural Sciences and Engineering Research Council of Canada (R.G. McLenaghan).

}


\begin{thebibliography}{99}

\bibitem{rPS}{S.Carloni, L.Fatibene, M.Ferraris, R.G. McLenaghan, P.Pinto,
 {\it Discrete Relativistic Positioning Systems},
GRG, {\bf 52}, (2), (2020)}

% Positioning puro
\bibitem{Coll1}{B. Coll, in: Proc. ERE-2000 Meeting on Reference Frames and Gravitomagnetism, eds. J. F. Pascual- Snchez, L. Flora, A. San Miguel, and F. Vicente (World Scientific, Singapore, 2001) 53.}

\bibitem{Blago}{M. Blago jevc, J. Garecki, F. W. Hehl, and Y. N. Obukhov, 
{\it Real null coframes in general relativity and GPS type coordinates}, Phys. Rev. D65 (2002) 044018.}

\bibitem{Coll2}{B.Coll, 
{\it Relativistic Positioning Systems}, in: Proc. Spanish Relativity Meeting ERE-2005, Oviedo (Spain); arXiv:gr-qc/0601110}

\bibitem{Coll3}{B.Coll, J.J.Ferrando, J.A.Morales, 
{\it Two-dimensional approach to relativistic positioning systems}, Phys. Rev. D73 (2006) 084017; arXiv:gr-qc/0602015}

\bibitem{Coll4}{B.Coll, J.M.Pozo, {\it Relativistic Positioning Systems: The Emission Coordinates}, Class. Quantum Grav.23 (2006) 7395; arXiv:gr-qc/0606044}

\bibitem{Coll5}{B.Coll, J.J.Ferrando, J.A.Morales, {\it Positioning with stationary emitters in a two-dimensional space-time}, Phys. Rev. D74 (2006) 104003; arXiv:gr-qc/0607037}

\bibitem{Tarantola}{A. Tarantola, L. Klimes, J. M. Pozo and B. Coll, {\it Gravimetry, Relativity, and the Global Navigation Satellite Systems}, arXiv:gr-qc/0905.3798}

\bibitem{Rey}{M. Lachieze-Rey, {\it The Covariance of GPS Coordinates and Frames}, Classical and Quantum Gravity 23.10 (2006), p. 3531. doi: 10.1088/0264- 9381/23/10/019.}






% theoretical positining

\bibitem{Gauss1845}{C.F. Gauss,
{\it Untersuchungen über Gegenstände der Höheren Geodäsie. Erste Abhandlung}, 
Abhandlungen der Königlichen Gesellschaft der Wissenschaften in Göttingen. Zweiter Band: 3–46.}


\bibitem{Bini}{D. Bini et al., {\it Emission versus Fermi Coordinates: Applications to Relativistic Positioning Systems}. Classical and Quantum Gravity 25.20 (2008), p. 205011. doi: 10.1088/0264-9381/25/20/205011.}

\bibitem{Rovelli1}{C. Rovelli, {\it GPS observables in general relativity}, Phys. Rev. D65 (2002) 044017; arXiv:gr- qc/0110003}

\bibitem{Schneider}{P.Schneider, J.Ehlers, E.E.Falco
{\it Graviational Lenses},  Springer-Verlag (Berlin, 1975) 514.}


% world function

\bibitem{Hadamard}{J. Hadamard {\it Lectures on Cauchy's Problem in Linear Partial Differential Equations}, Yale University Press (1923) 89}

\bibitem{Ruse1}{H.S. Ruse, {\it Taylor?s theorem in the tensor calculus}, Proc. London Math. Soc.32 (1931) 87}

\bibitem{Ruse2}{H.S.Ruse, {\it An absolute partial differential calculus}, Quart. J. Math. Oxford Ser. 2 (1931) 190}

\bibitem{Ruse3}{H.S.Ruse, A.G.Walker, T.J.Willmore. {\it Harmonic Spaces}, Ed. Cremonese Roma, (1961) 13}

\bibitem{Synge1}{J.L.Synge, A characteristic function in Riemannian space and its applications to the solution of geodesic triangles, Proc. London Math. Soc.32 (1931) 241.}

\bibitem{Yano}{K.Yano, Y.Muto, {\it Notes on the derivation of geodesics and the fundamental scalar in a Riemannian space}, Proc. Phys.-Math. Soc. Jap. 18 (1936) 142.}

\bibitem{Schouten}{J.A.Schouten, Ricci-Calculus. {\it An introduction to tensor analysis and its geometrical applications}, Springer-Verlag (1954) 382.}


\bibitem{Friedlander}{F.G.Friedlander,  
{\it The wave equation on a curved space-time},  Cambridge University Press (Cambridge, 1975) 17.}


\bibitem{Benenti}{S.Benenti, 
{\it Hamiltonian Structures and Generating Families}, Universitext (Springer-Verlag New York, 2011).}

\bibitem{Synge2}{J.L.Synge,  
{\it Relativity: The General Theory},  North-Holland (Amsterdam, 1960).}






\bibitem{EPS}{J.Ehlers, F.A.E.Pirani, A.Schild, {\it The Geometry of Free Fall and Light Propagation}, in: General Relativity, ed. L.O.‘Raifeartaigh (Clarendon, Oxford, 1972).}

\bibitem{OurEPS1}{M. Di Mauro, L. Fatibene, M.Ferraris, M.Francaviglia, {\it Further Extended Theories of Gravitation: Part I}, Int. J. Geom. Methods Mod. Phys. Volume: 7, Issue: 5 (2010), pp. 887-898; gr-qc/0911.2841}

\bibitem{OurEPS2}{ L. Fatibene, M. Ferraris, M. Francaviglia, S. Mercadante,
{\it Further Extended Theories of Gravitation: Part II},
Int. J. Geom. Meth. Mod. Phys. 7, (2010); %:899-906
arXiv:0911.2842 [gr-qc]
}




\bibitem{link}{Maple simulations can be found at {\tt http://www.fatibene.org/gps.html}}

\end{thebibliography}
\end{document}